\documentclass[preprint, 1p, showpacs, superscriptaddress, sort&compress, times]{elsarticle}
\usepackage{graphicx}
\usepackage{pifont, natbib, geometry,fleqn, graphicx, txfonts}
\usepackage{amsfonts}
\usepackage{amssymb}
\usepackage{amsmath}
\usepackage{epstopdf}
\usepackage{multirow}
\usepackage{xcolor}
\usepackage{hyperref}
\hypersetup{
	colorlinks=true,
	linkcolor=blue
}

\newcommand{\be}{\begin{equation}}
\newcommand{\ee}{\end{equation}}
\newcommand{\fL}{\mathcal{L}}
\newcommand{\SU}{\text{SU}}
\newcommand{\tU}{\text{U}}
\newcommand{\tO}{\text{O}}
\newcommand{\tr}{\text{tr}}
\newcommand{\fm}{\SU(n)/[\tU(1)]^{n-1}}
\newcommand{\CP}{\mathbb{C}\text{P}}
\usepackage{color}
\newcommand{\lcd}{\text{lcm}}
\usepackage{youngtab}
\newcommand{\ho}{\hspace{5mm} +\text{H.O.}}
\usepackage{tikz}
\usetikzlibrary{calc}
\definecolor{e1}{RGB}{255,0,0}
\definecolor{e6}{RGB}{255,165,0}
\definecolor{e4}{RGB}{255,255,0}
\definecolor{e3}{RGB}{0,255,0}
\definecolor{e2}{RGB}{0,0,255}
\definecolor{e7}{RGB}{75,0,130}
\definecolor{e5}{RGB}{238,130,238}
\definecolor{e8}{RGB}{0,191,255}
\newcommand{\cir}[1]{\tikz\draw[black, fill = #1, radius=3pt] (0,0) circle ;}%

\begin{document}

\title{Flag manifold sigma models from SU($n$) chains}
\date{\today}                                     

\author[ubc]{Kyle Wamer\corref{correspondingauthor}}
\author[ubc]{Ian Affleck}
\address[ubc]{Department of Physics and Astronomy and Stewart Blusson Quantum Matter Institute, University of British Columbia, 
Vancouver, B.C., Canada, V6T1Z1}

\cortext[correspondingauthor]{Corresponding author}

\begin{abstract}

One dimensional SU($n$) chains with the same irreducible representation $\mathcal{R}$ at each site are considered. We determine which $\mathcal{R}$ admit low-energy mappings to a $\fm$ flag manifold sigma model, and calculate the topological angles for such theories. Generically, these models will have fields with both linear and quadratic dispersion relations; for each $\mathcal{R}$, we determine how many fields of each dispersion type there are. Finally, for purely linearly-dispersing theories, we list the irreps that also possess a $\mathbb{Z}_n$ symmetry that acts transitively on the $\fm$ fields. Such SU($n$) chains have an 't Hooft anomaly in certain cases, allowing for a generalization of Haldane's conjecture to these novel representations. In particular, for even $n$ and for representations whose Young tableaux have two rows, of lengths $p_1$ and $p_2$ satisfying $p_1\not=p_2$, we predict a gapless ground state when $p_1+p_2$ is coprime with $n$. Otherwise,  we predict a gapped ground state that necessarily has spontaneously broken symmetry if $p_1+p_2$ is not a multiple of $n$.

\end{abstract}

\maketitle

\section{Introduction}

In 2017, Haldane's conjecture was generalized to SU(3) chains with a rank-$p$ symmetric representation on each site.\cite{LajkoNuclPhys2017} It was found that if $p$ is a multiple of 3, a finite energy gap exists above the ground state, while for all other values of $p$, gapless excitations exist. Since these rank$-p$ representations correspond to spin-$\frac{p}{2}$ in SU(2), this result is a quite natural extension of Haldane's original claim.\cite{HaldanePRL1983, HaldanePLA1983} For an extensive historical review of this subject, see \cite{LajkoNuclPhys2017}.

	In deriving the SU(3) result, the familiar correspondence between spin-$s$ antiferromagnet and O(3) sigma model with topological angle $\theta=2\pi s$ was necessarily generalized. This was first done by Bykov\cite{BYKOV2012100, bykov2013geometry}, and then repeated in \cite{LajkoNuclPhys2017, sun2019}. Now, the low energy physics of the SU(3) chain is captured by a SU(3)/[U(1)]${}^2$ flag manifold sigma model, with topological angles $\theta_1 = \frac{2\pi p }{3}$ and $\theta_2=\frac{4 \pi p}{3}$. And just as the original mapping to the O(3) model garnered interested from the larger theoretical physics community, so too has this SU(3) generalization: In the works \cite{Tanizaki:2018xto} and \cite{ohmori2019sigma}, SU($n$)/[U(1)]${}^{n-1}$ flag manifold models with additional discrete symmetries were analyzed in great detail. Of particular interest to those authors were the presence of 't Hooft anomalies in these theories, which indicate the presence of nontrivial physics at low energies: namely, spontaneous symmetry breaking or gapless excitations.

After the appearance of these two thoughtful papers on flag manifold sigma models and their anomalies, two further extensions of Haldane's conjecture followed. First, SU(3) chains with a self-conjugate representation at each site were shown to be described by the same SU(3)/[U(1)]${}^2$ sigma model, but without Lorentz-invariance, and with different topological angles.\cite{Wamer2019} In this case, a finite energy gap above the ground state was predicted in all cases, with spontaneously broken parity when $p$ is odd. The second extension considered again the rank$-p$ symmetric representations, but this time for generic SU($n$) chains. While these chains were already shown by Bykov to correspond to the (Lorentz-invariant) flag manifold sigma models considered by \cite{Tanizaki:2018xto} and \cite{ohmori2019sigma} for a fine-tuned set of coupling constants, in \cite{sun2019}, the renormalization group was used to show that this occurs in general, at low enough energies. By applying the 't Hooft anomaly conditions of these sigma models, as well as the results of \cite{Yao:2018kel}, the following SU($n$) Haldane conjecture was put forward: when $p$ and $n$ have no common divisor, gapless excitations are present above the ground state; otherwise, a finite energy gap exists. Recently, this result was also obtained in a different way, by considering how fractional instantons in the $\fm$ sigma model generate a mass gap.\cite{meron} When $p$ and $n$ are coprime, the fractional instantons destructively interfere, and the mass generating mechanism breaks down.

Based on this fruitful back-and-forth between SU($n$) chains and flag manifold sigma models, a more complete understanding of this correspondence is called for. What this entails is addressing the following question: What representations of SU($n$) chains give rise to SU($n$)/[U(1)]${}^{n-1}$ flag manifold sigma models? Once this has been answered, we should then further ask: What representations of SU($n$) chains give rise to those flag manifold sigma models that posses 't Hooft anomalies, and lend themselves to a Haldane-like prediction of gapless excitations in certain cases.

In this paper, we answer these two questions, ultimately classifying all SU($n$) chains that admit such a flag manifold sigma model mapping. In Section~\ref{sec:flag}, we introduce flag manifold sigma models, and discuss their various symmetry properties. In Section~\ref{sec:rep}, we review the representation theory of SU($n$) that is required for our analysis. In Section~\ref{sec:ham}, we introduce the SU($n$) chain Hamiltonian, and present the classification of its classical ground states and symmetries. Next, in Section~\ref{sec:disp}, we classify the topological angles and dispersion relations for generic SU($n$) chains. We will show that generically, both quadratic and linear dispersing modes are present in the corresponding field theories, which is a manifestation of the ferro- and antiferromagnetic order parameters that are jointly possible in these models. In a later work \cite{followup}, we will examine this intriguing phenomenon in greater detail, and explain how it may give rise to a fascinating hierarchy of flag manifold sigma models with novel 't Hooft anomalies. Finally, in Section~\ref{sec:result},  we summarize our calculations and present a new family of SU($n$) representations that give rise to linearly dispersing $\fm$ sigma models with 't Hooft anomalies, leading to a new generalization of Haldane's conjecture to SU($n$) chains.

\section{Flag Manifold Sigma Models} \label{sec:flag}

It is well known that the low energy physics of the SU(2) spin chain is described by the O(3) nonlinear sigma model.\cite{HaldanePRL1983, HaldanePLA1983} By sigma model, we mean a quantum field theory whose fields define a map from (Euclidean) space time to some curved manifold. In fact, the name O(3) sigma model is actually a misnomer, since in this case the field theory consists of a vector $\vec{n}\in\mathbb{R}^3$ that is constrained to live on the 2-sphere, which is not quite the Lie group O(3). Indeed, the relation that does hold is
\be \label{eq:0}
	S^2 = \tO(3)/\tO(2)
\ee
so that a more apt name would be the $S^2$ sigma model, or the O(3)/O(2) coset sigma model.\cite{altland2010condensed} To complicate the story further, one often introduces the complex field $\vec{\phi}\in\mathbb{C}^2$ through the relation
\be \label{eq:1}
	\vec{n} = \phi_\alpha^* \vec{\sigma}_{\alpha\beta} \phi_\beta,
\ee
which allows for a reformulation of the O(3) sigma model in terms of a normalized $\vec{\phi}$. Due to the presence of both $\vec{\phi}$ and $\vec{\phi}^*$ in the definition (\ref{eq:1}), this procedure also introduce a U(1) gauge symmetry, so that the underlying manifold that $\vec{\phi}$ lives on is 
\be \label{eq:2}
	\SU(2)/\tU(1) = \CP^1.
\ee
This establishes an equivalent field theory description of the antiferromagnet: the $\CP^1$ sigma model. While these technical subtleties may seem unnecessarily mathematical when analyzing an ordinary spin chain, they prove to be essential when promoting the symmetry group of the chain to SU($n$). This is because for $n>2$, the manifolds $\CP^{n-1}$ and $S^n$ are no longer equivalent, and so it begs the question: what is the appropriate generalization of the spin-chain/sigma model correspondence beyond SU(2)?

The answer to this question is not so simple, as was first realized by Affleck in \cite{Affleck:1984ar}, as it depends on the chosen representation of SU($n$) that occurs on each site of the chain. This fact will be explained in further detail in the following section. Instead, it is better to ask a slightly different question: for a specified sigma model, what is the appropriate generalization of the SU(2) spin chain? Over the years, this question has been partially answered for both $\CP^{n-1}$ models, and their Grassmann generalization, with symmetry group $\tU(n)/[ \tU(m)\times \tU(n-m)]$ (the case $m=1$ corresponds to $\CP^{n-1}$).\cite{AffleckSUn1988, Affleck1986, PhysRevB.42.4568} In this paper, we instead focus on a class of sigma models whose fields live on the flag manifold $\SU(n)/[\tU(1)]^{n-1}$, which is yet another generalization of the SU(2) case according to (\ref{eq:2}). We seek to find the complete set of SU($n$) chains that admit a mapping to these flag manifold sigma models (FMSMs). As mentioned in the introduction, these theories have attracted significant interest in recent years, culminating in a classification of their 't Hooft anomalies with various discrete symmetry groups \cite{Tanizaki:2018xto,ohmori2019sigma}. And while some mappings have already been identified \cite{BYKOV2012100, bykov2013geometry,LajkoNuclPhys2017, sun2019, Wamer2019}, an exhaustive list will provide a much larger testing ground for these theories, as well as serve to generalize the Haldane conjecture to a larger class of SU($n$) chains.

For concreteness, let us write down the Lagrangian for these flag manifold sigma models. It consists of $n$ orthonormalized fields $\vec{\phi}^\alpha \in \mathbb{C}^n$, which are each invariant under a U(1) gauge symmetry
\be
	\vec{\phi}^\alpha \mapsto e^{i\theta}\vec{\phi}^\alpha.
\ee
If we denote by $g_{\alpha\beta}$ the (symmetric) metric on the flag manifold, and $b_{\alpha\beta}$ the (antisymmetric) torsion, then the Lagrangian is\cite{ohmori2019sigma}
\be
	\fL = \sum_{\alpha,\beta=1}^n \left[ g_{\alpha\beta}\delta^{\mu\nu} + b_{\alpha\beta} \epsilon^{\mu\nu}\right] (\vec{\phi}^\alpha\cdot \partial_\mu \vec{\phi}^{\beta,*})(
	\vec{\phi}_\beta \cdot \partial_\nu \vec{\phi}^{\alpha,*})
	+ \fL_{\text{top}}
\ee
where $\fL_{\text{top}}$ is the topological term
\be
	\fL_{\text{top}} = \sum_{\alpha=1}^n \frac{\theta_\alpha }{2\pi} \epsilon^{\mu\nu} \partial_\mu \vec{\phi}^\alpha \cdot \partial_\nu \vec{\phi}^{\alpha,*}.
\ee
Only $n-1$ of the angles $\theta_\alpha$ are independent, since the theory is invariant under shifting all angles by the same amount. This reflects the fact that 
\be
	H_2(\SU(n)/[\tU(1)]^{n-1}) = \otimes_{k=1}^{n-1} \mathbb{Z}.
\ee
In fact, all of the topological angles may be removed using the combined shifts
\be
	\theta_\alpha \to \theta_\alpha + 2\pi b_\alpha \hspace{10mm}
	\text{ + } 
	\hspace{10mm}
	b_{\alpha\beta} \mapsto b_{\alpha\beta} -b_\alpha - b_\beta
\ee
but this hides the $2\pi$ periodicity of the $\theta_\alpha$. Finally, the shift $g_{\alpha\beta} \to g_{\alpha\beta} - c_\alpha - c_\beta$ introduces the familiar $\CP^{n-1}$ kinetic terms into the Lagrangian:
\be
	\fL \to \sum_{\alpha=1}^n c_\alpha \left[ |\partial_\mu \vec{\phi}^\alpha|^2 - |\vec{\phi}^\alpha \cdot \partial_\mu \vec{\phi}^\alpha|^2\right].
\ee
Based on this fact, it is useful to use think of the embedding $ \SU(n)/[\tU(1)]^{n-1} \hookrightarrow \otimes_{k=1}^n \CP^{n-1}$ and visualize the field content as a set of orthogonal $\CP^{n-1}$ fields, coupled through the metric and torsion terms.

In most cases, the tensors $g_{\alpha\beta}$ and $b_{\alpha\beta}$ will admit additional, discrete symmetries. For example, a sigma model that arises from an SU($n$) chain with a $d$-site unit cell in its classical ground state will posses a $\mathbb{Z}_d$ symmetry as a manifestation of the translation symmetry on the chain.

However, what is also true is that in most cases, the SU($n$) chain will not directly map to the above Lorentz-invariant sigma model. There are two reasons for this. The first is that the fields $\vec{\phi}^\alpha$ are not guaranteed to propagate with the same velocity. Indeed, for the symmetric representation SU($n$) chains, it was shown that only for a fine-tuned choice of SU($n$) chain coupling constants do these velocities become equal.\cite{sun2019} However, in the same paper, it was established that at low enough energies, all of the velocity differences flow to zero in the renormalization group sense. In this article, we will assume that this mechanism holds more generally, so that we may identify the various velocities of the $\CP^{n-1}$ fields. 

The second reason for Lorentz-non-invariance is more of a hinderance. It follows from a mismatch of terms arising from the coherent state path integral construction, ultimately leading to some of the $\vec{\phi}^\alpha$ having quadratic dispersion. In a later work\cite{followup}, we will discuss the consequences of this: in short, since Coleman's theorem does not apply in 1+1 dimensions to modes with quadratic dispersion, these quadratic modes may spontaneously order, resulting in true Goldstone modes with quadratic dispersion.\cite{Brauner_2010,Watanabe_2012, Watanabe:2014fva, beekman} These Goldstone bosons will couple to the remaining linear modes, which themselves form a $\SU(n')/[\tU(1)]^{n'-1}$ flag manifold sigma model with $n'<n$. If a subgroup of the translation symmetry acts transitively on the $n'$ linear fields, then it becomes possible for a novel 't Hooft anomaly (mixed with the $\mathbb{Z}_{n'}$ subgroup) to exist in such SU($n$) chains. Details of this mechanism will appear in \cite{followup}. In this paper, we avoid these complications by focusing only on theories with purely linearly-dispersing modes. To begin, we must first review the subject of SU($n$) representation theory.

\section{SU($n$) Representation Theory} \label{sec:rep}

Keeping in theme with the previous section, we begin with a review of SU(2) representation theory, and then outline how this generalizes to larger groups. A natural starting point is the three generators $\vec{S}^i$ of spin, which are used to write the nearest-neighbour Heisenberg interaction, $\vec{S}(i)\cdot\vec{S}(i+1)$. These generators obey the $\mathfrak{su}(2)$ Lie algebra
\be \label{eq:5}
	[S^i, S^j] = i\epsilon_{ijk}S^k,
\ee
and their associated SU(2) representation is completely specified by a single positive integer $p_1$, which can be found from the identity
\be \label{eq:4}
	\vec{S}(i)\cdot\vec{S}(i) = \frac{p_1}{4}(p_1+2)\mathbb{I}.
\ee
For physicists, we prefer the notation $s=\frac{1}{2} p_1$, and use the name spin-$s$ to refer to this representation. This relation (\ref{eq:4}) is a so-called Casimir constraint. In SU(2) it is the only one, but more generally there are $n-1$ such constraints, and together they ultimately dictate the target space manifold of our sigma models. Already in SU(2) this is apparent: in the limit of large representation (an assumption that we will always make), the commutator (\ref{eq:5}), together with the uncertainty relation
\be
	\Delta S^i \Delta S^j \sim | \langle [S^i, S^j]\rangle |,
\ee
allows for the operator $\vec{S}$ to be replace with a classical vector $\vec{n} \in \mathbb{R}^3$. The Casimir constraint (\ref{eq:4}) then restricts $\vec{n}$ to lie on the manifold $S^2$, leading to the ``O(3) sigma model'' description of the antiferromagnetic spin chain.

While this is the most familiar way of writing a spin chain, it will prove very useful to replace the vector $\vec{S}$ with a traceless matrix of operators, $S_{\alpha\beta}$. This follows from the fact that the number of generators of SU($n$) grows like $n^2$. Explicitly in SU(2), we define
\be \label{eq:su2}
	S = \begin{pmatrix} S^z & \frac{1}{2}(S^x -iS^y) \\
	\frac{1}{2}(S^x + iS^y) & -S^z \\
	\end{pmatrix}.
\ee
For all values of $n$, these matrices obey the commutation relations
\be
	[S_{\alpha\beta},S_{\mu\nu}] = \delta_{\alpha\mu}\delta_{\beta\nu}
	- \delta_{\alpha\nu}\delta_{\beta\mu}.
\ee
It is easily shown that the Heisenberg interaction can be rewritten in matrix form according to 
\be
	\vec{S}(i)\cdot\vec{S}(j) = \frac{1}{2}\tr[ S(i)S(i+1)]
\ee
and indeed this will be our starting point for constructing SU($n$) Hamiltonians in the next section. In the limit of large representation, the $2\times 2$ matrix in (\ref{eq:su2})  becomes a classical matrix, whose eigenvalues are entirely determined by the Casimir constraint (\ref{eq:4}): 
\be
	S = U^\dag \text{diag}(\lambda, -\lambda) U \hspace{10mm} \lambda^2 = \frac{p_1}{4}(p_1+2) 
\ee
The matrix $S$ now plays the role of $\vec{n}$, and since its eigenvalues are fixed, its target manifold is $\tU(2)/[\tU(1)]^2 = \SU(2)/\tU(1)$. A convenient parametrization of this space is in terms of the two orthonormal rows of $U$, each of which is invariant under a local U(1) rotation. This demonstrates the equivalent $\CP^1$ sigma model description of the antiferromagnet.

\begin{figure}[h]
\centering
\includegraphics[width =.9\textwidth]{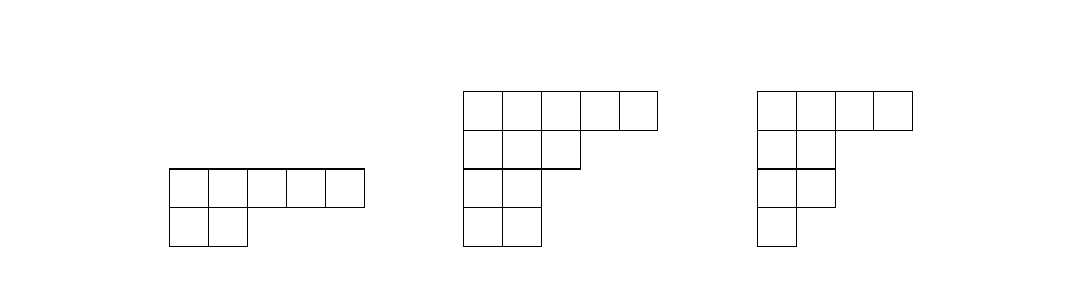}
\caption{Examples of Young tableaux in SU($n$). A diagram with $k$ nonzero rows corresponds to a representation in SU($n$) with $n\geq k+1$.}
\label{fig1}
\end{figure}

\begin{figure}[!h]
\centering
\includegraphics[width=.9\textwidth]{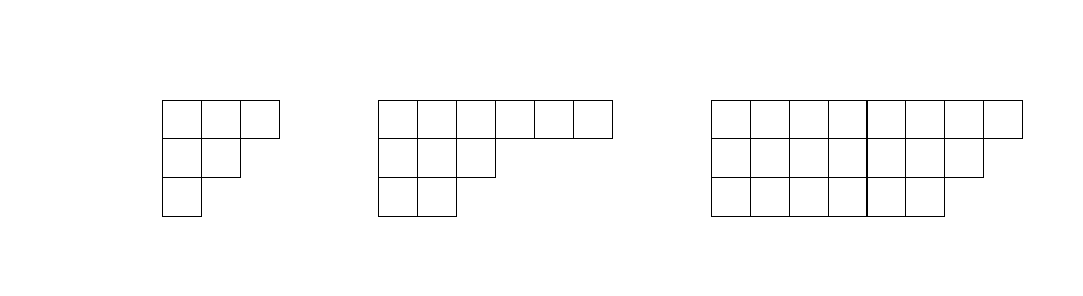}
\caption{Examples of Young tableaux that restrict $S$ to live in the $\SU(n)/[\tU(1)]^{n-1}$ flag manifold sigma model.}
\label{fig2}
\end{figure}

Now we repeat these steps for general $n$. We define a traceless $n\times n$ matrix $S_{\alpha\beta}$ whose entries correspond to the $n^2-1$ $\mathfrak{su}(n$) generators. Unlike SU(2), we must now specify more than one non-negative integer in order to label the representation. These integers are conveniently defined in terms of row lengths of a standard Young tableau. Indeed, the most general representation $[p_1,p_2,\cdots,p_{n-1}]$ of SU($n$) corresponds to a unique diagram of boxes arranged in $n-1$ rows, of lengths $p_1,p_2,\cdots, p_{n-1}$ respectively. The row lengths must satisfy $p_1\geq p_2\cdots \geq p_{n-1} \geq 0$. See Fig.~\ref{fig1} for some examples. In the limit $p_1\to \infty$, the matrix of operators is again replaced with a classical matrix $S$.\begin{footnote}{Since the quadratic Casimir $\tr[S^2]\to\infty$ when $p_1\to\infty$, it is sufficient to take this limit to obtain the classical limit; it is not necessary to also require $p_i\to\infty$ for $i>1$.}\end{footnote} Its eigenvalues are again completely determined, this time by $n-1$ distinct Casimir constraints 
\be
	\tr[ S^m] = C_m \mathbb{I} \hspace{5mm} m =2,3,\cdots, n.
\ee 
In terms of the row lengths $p_i$, the eigenvalues of $S$ are\cite{AffleckSUn1988}
\be
	\lambda_i = p_i - p \hspace{10mm} p: = \frac{1}{n}\sum_{i=1}^n p_i
\ee
where we've defined $p_n := 0$. Now, it becomes apparent how different representations of SU($n$) may lead to different types of sigma model. Indeed, the matrix $S$ is constrained to live on the manifold $\tU(n)/H$, where
\be \label{eq:9}
	H = \tU(m_1)\times \tU(m_2) \times\cdots \times \tU(m_k)
\ee 
is a product of $k$ unitary groups, one for each distinct value of $\lambda_i$, and $m_i$ is the degeneracy of each $\lambda_i$. Thus, it is possible to fix the target manifold of the matrices $S$ by choosing the appropriate representation of SU($n$) on each site. At this point, one might conclude that the only method to achieving our desired $\SU(n)/[\tU(1)]^{n-1}$ flag manifold sigma model is to ensure all the eigenvalues of $S$ are distinct. This amounts to considering representations whose Young tableaux have $n-1$ nonzero rows, each of a different length (see Fig.~\ref{fig2}). However, this is not the whole story, since multiple lattice sites must always be considered when deriving a sigma model description of an antiferromagnetic chain. As we will show in the following section, it is possible to work with representations that restrict $S$ to smaller manifolds, such as $\CP^{n-1}$, and then combine these degrees of freedom over consecutive sites of the chain to reproduce the larger flag manifold sigma model. This will also lead to additional discrete symmetries, as the translational invariance on the chain becomes a $\mathbb{Z}_d$ symmetry in the field theory, where $d$ is the size of the unit cell. This was the procedure used in \cite{LajkoNuclPhys2017} and \cite{sun2019}. It is worth emphasizing the difference in this approach from the original SU($n$) chains considered in  \cite{Affleck:1984ar}. In that paper, and the related ones that followed \cite{AffleckSUn1988, Affleck1986, PhysRevB.42.4568}, a desired sigma model was generated by identifying the representation $\mathcal{R}$ that directly restricts $S$ to the sigma model's full manifold, and then placing $\mathcal{R}$ and its conjugate $\overline{\mathcal{R}}$ on even and odd sites of the chain, respectively. In this work, we insist on having the same representation on each site; however, when our procedure is used to generate the flag manifold sigma models using a two-site unit cell, it will reduce to the older method for self-conjugate representations. This is precisely what occurred for the self-conjugate SU(3) chains considered in \cite{Wamer2019}.

\section{SU($n$) Chain Hamiltonians} \label{sec:ham}

In the previous section, we introduced the traceless matrices $S_{\alpha\beta}$ that contain the $n^2-1$ generators of SU($n$). These objects allow for us to write down the generalized Heisenberg interaction in terms of a trace:
\be
	\tr[S(i)S(j)].
\ee
In the limit of large representation, we replace $S$ with $U^\dag \text{diag}(\lambda_1,\cdots,\lambda_n) U$, so that this interaction becomes
\be \label{eq:10}
	\tr[S(i) S(j)] \to \sum_{\alpha,\beta,\gamma,\delta} \lambda_\alpha \lambda_\beta 
	|\vec{\phi}^{\alpha,*}(i)\cdot\vec{\phi}^\beta(j)|^2,
\ee
where we've defined 
\be
	U_{\alpha\beta}(i) = \phi^\alpha_\beta(i).
\ee
Since the $\vec{\phi}^\alpha$ are rows of a unitary matrix, they must be mutually orthonormal on the same site. Implicit in this expression is our assumption that the same representation occurs at each site of the chain. Since $\lambda_\alpha = p_\alpha - p$ for the representation with Young tableau row lengths $p_\alpha$, we opt to shift $S$ by $p\mathbb{I}$ to simplify our calculations (this shifts the interaction term by an overall constant). Having done this, the simplest SU($n$) chain Hamiltonian, namely the nearest-neighbour model, becomes
\be \label{eq:101}
	H = J \sum_i \sum_{\alpha,\beta=1}^{n-1}p_\alpha p_\beta |\vec{\phi}^{\alpha,*}(i)\cdot\vec{\phi}^\beta(i+1)|^2,\hspace{10mm} J>0.
\ee
Note that the sums over $\alpha$ and $\beta$ stop at $n-1$, since $p_n=0$ by definition. This nearest-neighbour model is the logical starting point for any SU($n$) generalization of the antiferromagnetic spin chain. However, in most cases, we will be required to consider Hamiltonians with longer range interactions if we hope to map to the $\fm$ flag manifold sigma model. As explained above, the manifold on which $S$ lies is dictated by the fixed representation on each site. Except in the special case when all of the row lengths $p_\alpha$ are distinct and nonzero, $S$ will be restricted to some subspace of the $\fm$ manifold. In order to reconstruct the complete flag manifold, we must couple the $S$ matrices from neighbouring sites together. In Sec~\ref{k=1}, we review how this works for the case of the symmetric representations, which were considered in detail in \cite{sun2019}. In that case, only a single row $p_1\not=0$, so that the corresponding manifold of $S$ is $\CP^{n-1}$. Since the complete $\fm$ flag manifold consists of $n$ orthogonally coupled such fields, one must add up to $(n-1)$-neighbour interactions in order to couple $n$ of these fields together. Instead, if one couples less than $n$ sites of the chain together, there will be leftover degrees of freedom, which manifest as local zero modes, ultimately prohibiting any field theory mapping. 

In the following subsections, we explain how this construction generalizes as we increase the number of nonzero $p_\alpha$. Loosely speaking, the number of nonzero rows $k$ in the representation will correspond to the number of fields $\vec{\phi}$ that exist at each site of the chain. We will then take $\lambda := \frac{n}{k}$ consecutive sites together to produce a mapping to the complete flag manifold. However, this isn't the whole story, since the $k$ fields on each site can still be locally rotated amongst one another. This is resolved by adding a $\lambda$-neighbour interaction that freezes out these additional degrees of freedom, which is essentially mimicking what happens when a representation $\mathcal{R}$ is coupled to its conjugate representation $\overline{\mathcal{R}}$.

Before proceeding, we must also mention what occurs when two rows $p_\alpha$ and $p_\beta$ have the same length. While this produces a factor of $\tU(2)$ in the quotient group $H$ just as would having a row of zero length, the result is fundamentally different. In both cases, there are spurious local degrees of freedom on each site (corresponding to rotating the $\vec{\phi}$ fields into each other); however, the trick of adding a $\lambda$-range interaction does not freeze this additional symmetry in the case of $p_\alpha=p_\beta \not=0$. This should become apparent below. As a result, for the most general representation of SU($n$), we do not expect that a mapping to the complete flag manifold sigma model exists, and so henceforth we restrict to the representations that satisfy $p_\alpha\not=p_\beta$ for all nonzero row lengths. Of course, other types of flag manifold sigma models can easily be constructed using such representations, but this is beyond the scope of this paper.

\subsection{Pictorial representation for classical ground states}

\begin{figure}[h]
\centering
\includegraphics[width = 0.9\textwidth]{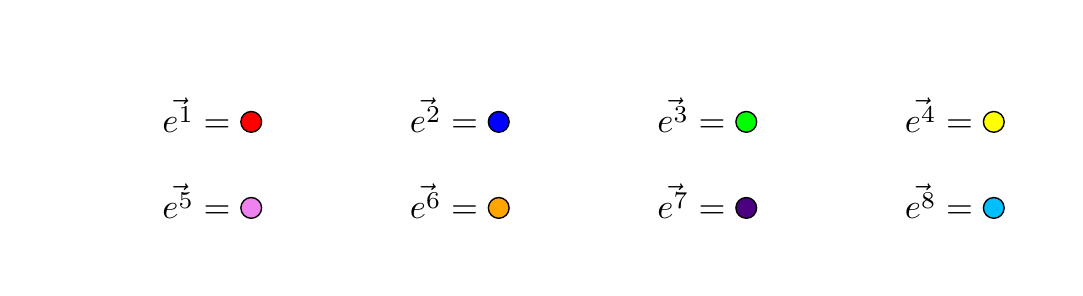}
\caption{colour dictionary for the first eight basis elements in $\mathbb{C}^n$. These coloured circles will be used to pictorially represent ground state states throughout.}
\label{fig:colour}
\end{figure}

In this subsection, we introduce some graphical notation that will aid in our classification of SU($n$) spin chains. According to (\ref{eq:10}), to each site of the chain we should assign a set of orthonormalized vectors $\vec{\phi}^\alpha$. We will make use of the standard orthonormal basis $\{\vec{e^\alpha} \}$ of $\mathbb{C}^n$, with
\be
	e^\alpha_\beta = \delta_{\alpha\beta}. 
\ee
We may use the same basis on each site of the chain, since any local change of basis transformation leaves the Hamiltonian invariant (and fortunately, no superpositions of states arise). Further, we will use coloured circles to represent the first few elements of this basis, in an effort to visually aid the reader. Our colour dictionary, for the first eight basis elements, can be found in Fig~\ref{fig:colour}.

\begin{figure}[h]
\centering
\includegraphics[width = 0.49\textwidth]{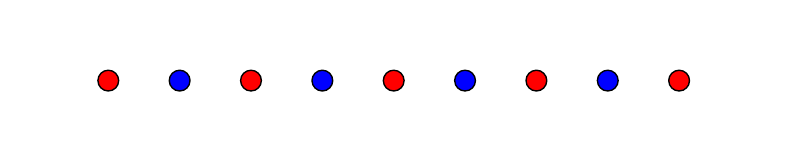}
\includegraphics[width = 0.49\textwidth]{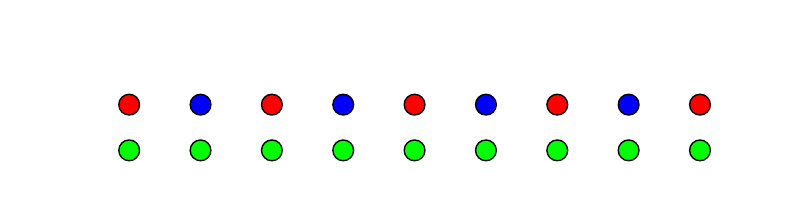}
\caption{Left: N\'eel state of the SU(2) antiferromagnet. Right: Classical ground state of the adjoint SU(3) chain, with $p_1=2$ and $p_2=1$.}
\label{eq:neel}
\end{figure}

When drawing a classical ground state, we will arrange the same-site vectors into a single column, and use a white space to separate neighbouring chain sites. For example, the N\'eel state of the SU(2) antiferromagnet is given in Fig~\ref{eq:neel} left, while a classical ground state of the adjoint SU(3) chain is given in Fig~\ref{eq:neel} right. This will be demonstrated below. The self-conjugate ground state is slightly different than the ones discussed in \cite{Wamer2019}, where a different convention for the matrices $S$ was chosen. 

\begin{figure}[ht]
\centering
\includegraphics[width=.99\textwidth]{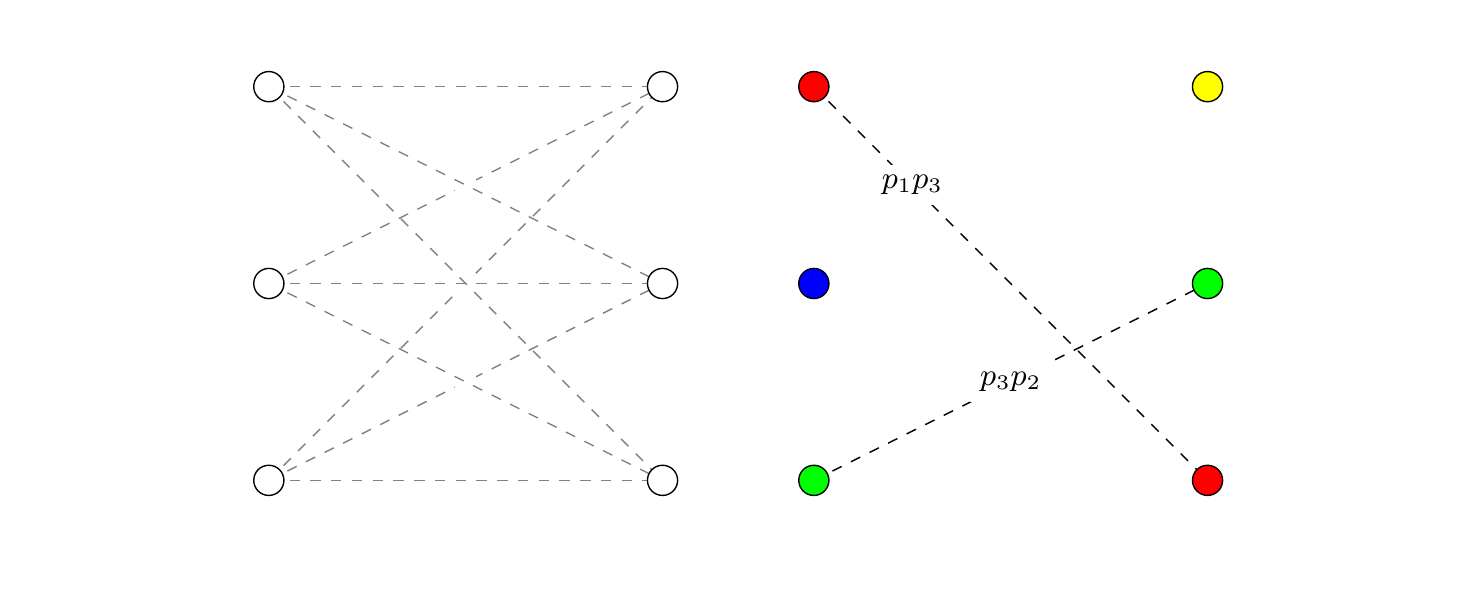}
\caption{Left: Imaginary bonds between two sets of vectors $\vec{\phi}^\alpha$ and $\vec{\phi}^\beta$ at sites $i$ and $j$ on the chain. Right: The energy cost of each bond is $p_{\alpha_0} p_{\beta_0}$ whenever two nodes have the same colour; $\alpha_0$ and $\beta_0$ are the column positions of these nodes. }
\label{figx}
\end{figure}

The benefit of these ground state pictures is that it makes it easy to read off the energy cost of a term $\tr[S(i)S(j)]$. Indeed, since each colour corresponds to a standard unit vector $\vec{e}^\alpha$, we have according to (\ref{eq:101}), 
\be \label{eq:key}
	\tr[S(i)S(j)] = \sum_{\alpha,\beta} p_\alpha p_\beta |\vec\phi^{\alpha,*}(i)\cdot\vec{\phi}^{\beta}(j)|^2.
\ee
The right hand side of this expression vanishes unless one of the complex unit vectors (i.e. one of the colours) at site $i$ equals one of the complex unit vectors at site $j$. In this case, the RHS equals $p_{\alpha_0} p_{\beta_0}$, where $\alpha_0$ and $\beta_0$ are the respective positions of the unit vector/colour in column $i$ and column $j$. To visualize this, it is useful to imagine bonds between all of the circles of the two columns, as in Fig~(\ref{figx}). These bonds are inactive (meaning zero energy cost), unless two nodes are the same colour. For example, the N\'eel state in (\ref{eq:neel}, left) has an energy cost of zero per site (recall we have shifted the $S_{\alpha\beta}$ matrices by a constant), while the classical ground state of the adjoint chain (\ref{eq:neel}, right) has energy cost of $p_2^2$ per site.

With this notation in place, we can now begin our classification of SU($n$) chains with representations that have $p_{\alpha}\not=p_{\beta}$ for all nonzero row lengths $p_{\alpha}$. Over the next few subsections, we subdivide this task into various cases according to how many nonzero rows are present in the representation. Throughout, we use the letter $k$ to refer to this number, and also define $\lambda = \lfloor \frac{n}{k} \rfloor$.

\subsection{Case 1: $k=1$.} \label{k=1}

As a warm-up to the more complicated representations below, we begin with reviewing what occurs for the symmetric representations of SU($n$), with Young tableaux that have a single row of length $p_1$. This case is discussed in more detail in \cite{LajkoNuclPhys2017} for SU(3), and in \cite{sun2019} for general $n$.

We start by considering a nearest-neighbour SU($n$) Heisenberg Hamiltonian, and list its classical ground states. According to (\ref{eq:key}), any configuration that has no energy cost per bond will be a classical ground state. Since $k=1$, and only a single node is present at each site, the N\'eel state (\ref{eq:neel}, left) is such an example. However, for $n>2$, the basis at each site is larger than 2 (i.e. there are other colours available), and this leads to an infinite number of other ground states. Indeed, the ground state 
\be
	\cir{e1} \hspace{5mm}\cir{e2}\hspace{5mm}\cir{e1}\hspace{5mm}\cir{e3}\hspace{5mm}\cir{e1}\hspace{5mm}\cir{e2}\hspace{5mm}\cir{e1}\hspace{5mm},
\ee
which exists for $n>2$, is related by a zero-energy transformation to the N\'eel state. This local degree of freedom is an example of a zero-energy mode, and destabilizes any candidate ground state above which we would like to derive a quantum field theory. As a result, the nearest-neighbour Hamiltonian must be modified if we would like to proceed. Since it is believed that longer-range interactions may be dynamically generated from the nearest-neighbour model \cite{PhysRevX.2.041013}, we add further-neighbour interactions to realize a stable ground state. Since there are $n$ possible colours, we require up to $(n-1)$-neighbour interactions, all of which are taken to be antiferromagnetic, in order to remove the zero modes. For example, in SU(5), with interactions up to 4th neighbour, one such ground state is
\be \label{eq:sym1}
	\cir{e1} \hspace{5mm}\cir{e2}\hspace{5mm}\cir{e3}\hspace{5mm}\cir{e4}\hspace{5mm}\cir{e5}\hspace{5mm}\cir{e1}\hspace{5mm}\cir{e2}\hspace{5mm} \cir{e3} \hspace{5mm} \cir{e4} \hspace{5mm}\cir{e5} \hspace{5mm}.
\ee
While this large number of interaction terms may seem contrived, there is second reason why one should consider adding them. Arguably, it is the simplest way to restrict to classical ground states that have a $\mathbb{Z}_n$ symmetry, which is to be expected for the symmetric chains, since this is a feature of the integrable SU($n$) chains, that correspond to $p=1$. 

In \cite{sun2019}, it was shown in great detail how the $\fm$ sigma model arises as the low energy description of this longer-range Hamiltonians. While the on-site matrix $S$ lies in in $\CP^{n-1}$ (as was explained in Section \ref{sec:flag}), by coupling $n$ neighbouring sites together, our underlying degrees of freedom are actually $n$ orthogonally coupled $\CP^{n-1}$ fields, which is equivalent to $\fm$. In the more general representations below, we will see a similar pattern: collections of $\CP^{n-1}$ fields from neighbouring sites will become orthogonally coupled, ultimately leading to the flag manifold sigma model that we desire.

\subsection{Case 2: $k=n-1$.} \label{sub:case2}

We now graduate to the second class of representations, which have Young tableaux with $n-1$ nonzero rows, and are arguably simpler than the symmetric representations considered above. Since the on-site representation of the $S$ matrix already corresponds to the manifold $\fm$, a nearest-neighbour Heisenberg interaction is sufficient to derive the associated sigma model. Let us first demonstrate this in SU(3). The interaction term
\be \label{eq:key}
	\tr[S(i)S(i+1)] = \sum_{\alpha,\beta=1}^2 p_\alpha p_\beta |\vec\phi^{\alpha,*}(i)\cdot\vec{\phi}^\beta(i+1)|^2
\ee
is never zero for two adjacent sites, which requires choosing the colour for four nodes. Since $p_1>p_2$, the minimum is $p_2^2$, which is achieved when the two same-colour nodes are in the second position of the column. Thus, a typical ground state in SU(3) looks like \newpage
\be
	\cir{e1} \hspace{5mm}\cir{e2}\hspace{5mm}\cir{e1}\hspace{5mm}\cir{e2}\hspace{5mm}\cir{e1}\hspace{5mm}\cir{e2}\hspace{5mm}\cir{e1}\hspace{5mm}
\ee
\vspace{-5mm}
\[
	\cir{e3} \hspace{5mm}\cir{e3}\hspace{5mm}\cir{e3}\hspace{5mm}\cir{e3}\hspace{5mm}\cir{e3}\hspace{5mm}\cir{e3}\hspace{5mm}\cir{e3}\hspace{5mm}
\]
which is precisely what we drew above for the adjoint SU(3) chain (which corresponds to the case $p_1=2, p_2=1$). Note that no local transformations exist that cost zero energy: all of the $p_2$ nodes must stay the same colour in order to minimize the $\tr[S(i)S(i+1)]$ term, and the remaining two colours behave just like the N\'eel state of SU(2). 

In SU(4), the $\tr[S(i)S(i+1)]$ requires the introduction of six coloured nodes. Using the inequality $p_2^2+p_3^2 \geq 2p_2p_3$, we see that the ground states have the following form:
\be
	\cir{e1} \hspace{5mm}\cir{e2}\hspace{5mm}\cir{e1}\hspace{5mm}\cir{e2}\hspace{5mm}\cir{e1}\hspace{5mm}\cir{e2}\hspace{5mm}\cir{e1}\hspace{5mm}
\ee
\vspace{-7mm}
\[
	\cir{e3} \hspace{5mm}\cir{e4}\hspace{5mm}\cir{e3}\hspace{5mm}\cir{e4}\hspace{5mm}\cir{e3}\hspace{5mm}\cir{e4}\hspace{5mm}\cir{e3}\hspace{5mm}
\]
\vspace{-6mm}
\[
	\cir{e4} \hspace{5mm}\cir{e3}\hspace{5mm}\cir{e4}\hspace{5mm}\cir{e3}\hspace{5mm}\cir{e4}\hspace{5mm}\cir{e3}\hspace{5mm}\cir{e4}\hspace{5mm}
\]
This pattern extends to general $n$: the first row of nodes establishes a N\'eel-like state, while the remaining $n-2$ rows have a ``reverse-ordered'' pattern: the colour ordering along a column switches direction between even and odd sites. In \ref{app:order}, we prove that these states indeed minimize the Hamiltonian. Here is an example ground state in SU(5):
\be \label{eq:su51}
	\cir{e1} \hspace{5mm}\cir{e2}\hspace{5mm}\cir{e1}\hspace{5mm}\cir{e2}\hspace{5mm}\cir{e1}\hspace{5mm}\cir{e2}\hspace{5mm}\cir{e1}\hspace{5mm}
\ee
\vspace{-7mm}
\[
	\cir{e3} \hspace{5mm}\cir{e4}\hspace{5mm}\cir{e3}\hspace{5mm}\cir{e4}\hspace{5mm}\cir{e3}\hspace{5mm}\cir{e4}\hspace{5mm}\cir{e3}\hspace{5mm}
\]
\vspace{-6mm}
\[
	\cir{e5} \hspace{5mm}\cir{e5}\hspace{5mm}\cir{e5}\hspace{5mm}\cir{e5}\hspace{5mm}\cir{e5}\hspace{5mm}\cir{e5}\hspace{5mm}\cir{e5}\hspace{5mm}
\]
\vspace{-6mm}
\[
	\cir{e4} \hspace{5mm}\cir{e3}\hspace{5mm}\cir{e4}\hspace{5mm}\cir{e3}\hspace{5mm}\cir{e4}\hspace{5mm}\cir{e3}\hspace{5mm}\cir{e4}\hspace{5mm}
\]
For these representations, the unit cell is always 2 sites in length, which leads to a $\mathbb{Z}_2$ translation symmetry in the sigma model.


\subsection{Case 3: $n = \lambda k$}

In this case, the matrix $S$ at each site of the chain lies neither in $\CP^{n-1}$ nor $\fm$. While it would be straightforward to derive other families of flag manifold sigma models from these representations, we are only interested in $\fm$. Thus, some care must be taken in order to realize the appropriate degrees of freedom. 

As before, we begin with an example, this time with $k=2$ in SU(4). This requires choosing four colours for four nodes in order to minimize the $\tr[S(i)S(i+1)]$ term, which is easily done. For example:
\be \label{eq:form}
	\cir{e1} \hspace{5mm}\cir{e2}\hspace{5mm}\cir{e1}\hspace{5mm}\cir{e2}\hspace{5mm}\cir{e1}\hspace{5mm}\cir{e2}\hspace{5mm}\cir{e1}\hspace{5mm}
\ee
\vspace{-6mm}
\[
	\cir{e3} \hspace{5mm}\cir{e4}\hspace{5mm}\cir{e3}\hspace{5mm}\cir{e4}\hspace{5mm}\cir{e3}\hspace{5mm}\cir{e4}\hspace{5mm}\cir{e3}\hspace{5mm}
\]
However, such a configuration does not lead to the manifold $\fm$, because the four colours do not behave like four orthogonal $\CP^{3}$ fields. Indeed, at each site, we may additionally rotate the two colours into each other at no energy cost, which corresponds to another type of zero mode. In order to achieve the correct flag manifold, we ``freeze out'' these additional degrees of freedom by adding a weaker second-neighbour interaction, $\tr[S(i)S(i+2)]$. The effect of this term is to invoke a ``reverse ordering'' pattern between sites and their second neighbours: the new ground state is

\noindent \begin{minipage}{\textwidth}
\be \label{eq:z4}
	\cir{e1} \hspace{5mm}\cir{e2}\hspace{5mm}\cir{e3}\hspace{5mm}\cir{e4}\hspace{5mm}\cir{e1}\hspace{5mm}\cir{e2}\hspace{5mm}\cir{e3}\hspace{5mm}
\ee
\vspace{-5mm}
\[
	\cir{e3} \hspace{5mm}\cir{e4}\hspace{5mm}\cir{e1}\hspace{5mm}\cir{e2}\hspace{5mm}\cir{e3}\hspace{5mm}\cir{e4}\hspace{5mm}\cir{e1}\hspace{5mm}
\]
\vspace{5mm}
\end{minipage}
The fact that this ground state minimizes the combined $J_1\tr[S(i)S(i+1)] + J_2\tr[S(i+1)S(i+2)$ term (for antiferromagnetic couplings $J_1 \gg J_2$) follows from the identity $p_1^2+p_2^2 \geq 2p_1p_2$. In a sense, this second-neighbour interaction generates the same behavior that we saw in the previous case of $k=n-1$: The nearest-neighbour term partitions the colours into subsets, and the second-neighbour term reverse-orders these subsets, effectively breaking the additional on-site rotation symmetry between colours. In the $k=n-1$ case, both of these steps are achieved by the same interaction term: first the colours are partitioned into 3 sets:  e.g. $\{\cir{e1}\}, \{\cir{e2}\}, \{\cir{e3}, \cir{e4}, \cir{e5}\}$, and then each set is reverse ordered compared to the previous time it occurred. It will turn out that this reverse ordering is a generic feature of all the representations that we consider.

As a next step, we extend from 4 to general even $n$, and consider $k=\frac{n}{2}$. A nearest-neighbour interaction will again serve to partition the colours into two sets, leaving a local rotation symmetry among the $k$ colours on each site. In order to freeze out these degrees of freedom, we again add a second-neighbour interaction, which reverse orders each set. For example, in SU(6), we have
\be  \label{eq:nk1}
	\cir{e1} \hspace{5mm}\cir{e4}\hspace{5mm}\cir{e3}\hspace{5mm}\cir{e6}\hspace{5mm}\cir{e1}\hspace{5mm}\cir{e4}\hspace{5mm}\cir{e3}\hspace{5mm}\cir{e6} \hspace{5mm}
\ee
\vspace{-6mm}
\[
	\cir{e2} \hspace{5mm}\cir{e5}\hspace{5mm}\cir{e2}\hspace{5mm}\cir{e5}\hspace{5mm}\cir{e2}\hspace{5mm}\cir{e5}\hspace{5mm}\cir{e2}\hspace{5mm} \cir{e5} \hspace{5mm} 
\]
\vspace{-6mm}
\[
	\cir{e3} \hspace{5mm}\cir{e6}\hspace{5mm}\cir{e1}\hspace{5mm}\cir{e4}\hspace{5mm}\cir{e3}\hspace{5mm}\cir{e6}\hspace{5mm}\cir{e1}\hspace{5mm}\cir{e4}\hspace{5mm}
\]
Clearly, the ground state will always have a 4-site unit cell for $k=\frac{n}{2}$.

Now, when $k<\frac{n}{2}$, the full set of colours is no longer used up when the nodes on two neighbouring sites are filled. As a result, additional zero modes are present that cannot be removed by reverse ordering the colours within a set. To resolve this, we first add up to $(\lambda-1)$-neighbour interactions (always with antiferromagnetic couplings), to properly partition the full set of $n$ colours into $\lambda$ sets of $k$ elements ($\lambda := \frac{n}{k}$). Then, we add a weaker $\lambda$-neighbour interaction which serves to reverse order within each set of the partition. For example, in SU(6) with $k=2$, the Hamiltonian we should consider is
\be \label{eq:nk2}
	H = \sum_i \Big( J_1 \tr[S(i)S(i+1)] +J_2 \tr[S(i)S(i+2)] + J_3 \tr[S(i)S(i+3)]\Big)
\ee
with $J_1>J_2 \gg J_3 >0$, which has, for example, the following ground state 
\be 
	\cir{e1} \hspace{5mm}\cir{e3}\hspace{5mm}\cir{e5}\hspace{5mm}\cir{e2}\hspace{5mm}\cir{e4}\hspace{5mm}\cir{e6}\hspace{5mm}\cir{e1}\hspace{5mm}\cir{e3} \hspace{5mm} \cir{e5} \hspace{5mm}
\ee
\vspace{-6mm}
\[
	\cir{e2} \hspace{5mm}\cir{e4}\hspace{5mm}\cir{e6}\hspace{5mm}\cir{e1}\hspace{5mm}\cir{e3}\hspace{5mm}\cir{e5}\hspace{5mm}\cir{e2}\hspace{5mm} \cir{e4} \hspace{5mm}\cir{e6} \hspace{5mm}
\]
The $J_1$ and $J_2$ terms serve to partition the colours into three sets: $\{\cir{e1}, \cir{e2} \}, \{ \cir{e3}, \cir{e4}\}, \{\cir{e5},\cir{e6}\}$, and the $J_3$ terms serve to reverse order within each of these three sets. Based off of this example, we can see that the unit-cell has size $2\lambda$ for these representations.

\subsection{Case 4: $n=\lambda k + c $}

Finally, we consider all remaining values of $k$. Let $c=n \text{ mod } k$, so that $n=\lambda k + c$ for some $\lambda \in \mathbb{Z}$. As in the previous case of $n=k\lambda$, local zero modes will be present unless sufficiently long range interactions are included to use up all of the available colours. We add up to $\lambda$-neighbour terms, which partitions the colours into $\lambda$ sets of $k$ elements, and one set of $c$ elements. Briefly, we return to the notation $\vec{e}^\alpha$ for the basis vectors instead of coloured nodes. Then, a possible partitioning of the colours is:
\be
	\{ e^1, e^2, \cdots, e^k \}, \{ e^{k+1},e^{k+2},\cdots, e^{2k}\},\cdots, \{e^{(\lambda-1)k + 1}, e^{(\lambda-1)k +2}, \cdots, e^{\lambda k} \},
	\{e^{\lambda k+1}, \cdots, e^{\lambda k +c}\}.
\ee
Now, in order to minimize the interaction term $\tr[S(i)S(i+\lambda)]$, the remaining nodes on the $(\lambda+1)$th site will be the reverse ordered set $\{e^k, e^{k-1},\cdots, e^{c+1}\}$. For example, in SU(7) with $k=3$,  three consecutive sites may look like
	\be 
	\cir{e1} \hspace{5mm}\cir{e4}\hspace{5mm}\cir{e7}\hspace{5mm}
\ee
\vspace{-7mm}
\[
	\cir{e2} \hspace{5mm}\cir{e5}\hspace{5mm}\cir{e3}\hspace{5mm}
\]
\vspace{-6mm}
\[
	\cir{e3} \hspace{5mm}\cir{e6}\hspace{5mm}\cir{e2}\hspace{5mm}
\]
 The nodes of the next site (which is a ($\lambda+1$)th neighbour), will then begin to be filled with the remaining $\{e^1,\cdots, e^c\}$ colours from the first site. In our present SU(7) example, this looks like: 
\be  \label{eq:su7}
	\cir{e1} \hspace{5mm}\cir{e4}\hspace{5mm}\cir{e7}\hspace{5mm}
	\cir{e1} \hspace{5mm} \cir{e4} \hspace{5mm} \cir{e7} \hspace{5mm}
\ee
\vspace{-7mm}
\[
	\cir{e2} \hspace{5mm}\cir{e5}\hspace{5mm}\cir{e3}\hspace{5mm}
	 \cir{e6}  \hspace{5mm} \cir{e2} \hspace{5mm} \cir{e5} \hspace{5mm}
\]
\vspace{-6mm}
\[
	\cir{e3} \hspace{5mm}\cir{e6}\hspace{5mm}\cir{e2}\hspace{5mm}
	\cir{e5} \hspace{5mm} \cir{e3} \hspace{5mm} \cir{e6} \hspace{5mm}
\]
 Since $c=1$ in this example, the drawn ground state is stable. However for $c>1$, there will still be zero modes associated with rotating among the set $\{e^1,\cdots, e^k\}$. Thus, an additional $(\lambda+1)$-neighbour interaction must also be added! The following ground state for SU(5) with $k=3$ demonstrates this:
 \be  \label{eq:su5}
	\cir{e1} \hspace{5mm}\cir{e4}\hspace{5mm}\cir{e2}\hspace{5mm}
	\cir{e5} \hspace{5mm} 
\ee
\vspace{-6mm}
\[
	\cir{e2} \hspace{5mm}\cir{e5}\hspace{5mm}\cir{e1}\hspace{5mm} \cir{e4}  \hspace{5mm} 
\]
\vspace{-6mm}
 \[
	\cir{e3} \hspace{5mm}\cir{e3}\hspace{5mm}\cir{e3}\hspace{5mm}
	\cir{e3} \hspace{5mm}
\]

Thus, we are led to the following conclusion for this class of representations: If $c=1$, our Hamiltonians should contain up to $\lambda$-neighbour interactions, and if $c>1$, we should also add an additional $(\lambda+1)$-neighbour interaction term.

\begin{figure}[!h]
\centering
\includegraphics[width=.8\textwidth]{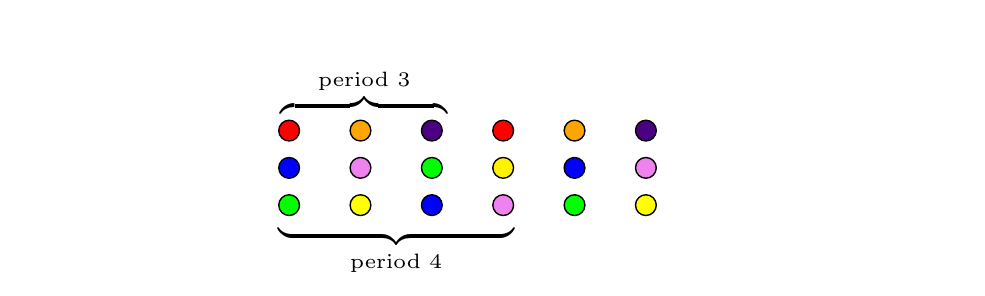}
\caption{Potential ground state of an SU(7) chain. Since the first row has 3-site periodicity, and the remaining rows have 4-site periodicity, the unit cell is 12 sites in length.} 
\label{eq:pattern}
\end{figure}

Using the emerging patterns in the previous examples as a guide, we may now determine the unit-cell size for the most general representation. This quantity is very important, as it determines the translation group symmetry that is present in the flag manifold sigma model. Note that in both (\ref{eq:su7}) and (\ref{eq:su5}), there are two competing types of order among the coloured nodes. The first $c$ rows exhibit one type of order, which has periodicity $\lambda+1$ when $c=1$, and $2(\lambda+1)$ otherwise. Meanwhile, the remaining $k-c$ rows have a periodicity $2\lambda$ for all $c$ except $c=k-1$, in which case the periodicity is $\lambda$. In order to determine the overall unit-cell length, we must find the least common multiple of these two periodicities. For example, in our SU(7) example, we see that the unit cell will have length 12, leading to a $\mathbb{Z}_{12}$ symmetry in the field theory. See Fig~\ref{eq:pattern}.

\subsection{Summary of Classification}

In Table~\ref{result}, we summarize our results from the previous subsections. In the first column, we specify the number of nonzero rows in the Young tableau, $k$, and the integer $c := n \text{ mod } k$. In the second column, we write down the longest-range interaction that must be included in the Hamiltonian in order to eliminate any local zero modes. As always, it is understood that each interaction term is $J_\alpha \tr[S(i)S(i+\alpha)]$ for some coupling $J_\alpha>0$, and that $J_{\alpha}> J_{\beta}$ for $\alpha<\beta$. Finally, in the third column, we specify the order $d$ of the translation group $\mathbb{Z}_d$ that acts on the corresponding flag manifold sigma model. This order equals the size of the unit-cell in the classical ground states of the Hamiltonian.

\begin{center}
\begin{table}[h]
\centering
\begin{tabular}{l | c | c}
Representation & Longest Interaction & Translation Group Order\\
\hline \hline 
$k=1$ & $J_{n-1} \tr[S(i)S(i +n-1)]$ & $n$ \\ \hline
$k=n-1$ & $J_1 \tr[S(i)S(i+1)]$ & $2$ \\ \hline
$k= \frac{n}{\lambda}, \hspace{15mm} \lambda < n$ & $J_\lambda \tr[S(i)S(i+\lambda)] $& $2\lambda$ \\
\hline
$n=2\lambda+1, \hspace{10mm}k=2$ & $J_\lambda\tr[S(i)S(i+\lambda)]$ & $\lambda(\lambda+1)$ \\ \hline
$ n = k\lambda + 1, \hspace{5mm} \lambda>1, k>2 $ & $J_\lambda \tr[S(i)S(i+\lambda)]$ & $\lcd[2\lambda,(\lambda+1)]$\\ \hline
$n = k\lambda + c, \hspace{5mm} c\not= 1, k-1$ & $J_{\lambda+1}\tr[S(i)S(i+\lambda+1)]$ & $2\lambda(\lambda+1)$\\ \hline
$n = k\lambda +(k- 1)$& $J_{\lambda+1}\tr[S(i)S(i+\lambda+1)]$ & $\lcd[\lambda,2(\lambda+1)]$\\ \hline
\end{tabular}
\caption{Classification results of all SU($n$) representations satisfying $p_{\alpha}\not=p_\beta$ for all nonzero $p_\alpha$. We use the notation $\lcd[a,b]$ to denote the least common multiple of $a$ and $b$.}
\label{result}
\end{table}
\end{center}

In the final column of the table, the following identities are useful:
\be
	\lcd[2\lambda,(\lambda+1)] = \begin{cases} \lambda(\lambda+1) & \lambda \text{ is odd } \\
	2\lambda(\lambda+1) & \lambda \text{ is even } \\
	\end{cases}
\ee
\be
	\lcd[\lambda,2(\lambda+1)] = \begin{cases} 2\lambda(\lambda+1) & \lambda \text{ is odd } \\
	\lambda(\lambda+1) & \lambda \text{ is even } \\
	\end{cases}
\ee
In the following section, we build on this classification, and determine the dispersion relations and topological angles in each class of representation.

\section{Dispersion Relations and Topological Angles} \label{sec:disp}

Now that we've determined the appropriate Hamiltonians of the most general SU($n$) chains admitting a $\fm$ sigma model description, we now turn to the field theory mapping itself. Of course, a detailed derivation for each Hamiltonian would be a very tedious undertaking, and we do not pursue this here. Instead, we focus on particular features, namely the topological angles and possible dispersion relations that exist in these theories, and refrain from determining the precise coupling constants and velocities as a function of the interaction strengths $\{J_\alpha\}$. In the following section, we will explain how these pieces of information, combined with a set of reasonable assumptions, will allow us to make predictions about the ground state behaviour of certain SU($n$) chains.

\subsection{Review of the Coherent State Path Integral Construction}

In order to construct a mapping from Hamiltonian to sigma model, we use coherent states to generate a path integral of the ground state fluctuations. These coherent states are constructed as follows. For a representation with nonzero Young tableau rows $p_1,\cdots, p_k$, we introduce $k$ orthonormal fields $\vec{\phi}^\alpha \in \mathbb{C}^n$, and $k$ $n$-component creation operators $\vec{a}^{\alpha,\dag}$. We then define\cite{Mathur_2010, mathur2}
\be
	|\Phi\rangle := \sum_{\alpha=1}^k [\vec \phi^\alpha \cdot \vec{a}^\alpha]^{p_\alpha} |0\rangle.
\ee
These are the coherent states of SU($n$), and in order to construct a path integral, we insert them between thin time slices of the partition function:
\be
	\langle \Phi(\tau_i) | e^{-H\delta \tau} |\Phi(\tau_i +\delta \tau) \rangle
	= \langle \Phi(\tau_i) | \Phi(\tau_i+\delta\tau) \rangle e^{-H\delta \tau}.
\ee
The right hand side can be approximated using 
\be
	\langle \Phi(\tau_i ) | \Phi(\tau_i+\delta\tau)\rangle 
	\approx \sum_{\alpha=1}^k\left[ 1 + \vec{\phi}^{\alpha,*}\cdot \partial_\tau \vec{\phi}^\alpha \right]^{p_\alpha},
\ee
which follows from 
\be
	\langle \Phi(\tau) |\Phi(\tau')\rangle = \sum_{\alpha=1}^k (\vec{\phi}^{\alpha,*}(\tau)\cdot \vec{\phi}^\alpha(\tau'))^{p_\alpha}.
\ee
By taking the product over all time slices $\tau_i$, we can then reexponentiate according to 
\be
	\prod_i\langle \Phi(\tau_i ) | \Phi(\tau_i+\delta\tau)\rangle 
	= \exp\sum_i \log \langle \Phi(\tau_i ) | \Phi(\tau_i+\delta\tau)\rangle 
	\approx \exp \sum_i  \sum_\alpha p_\alpha \vec \phi^{\alpha,*}\cdot \partial_\tau \vec{\phi}^\alpha.
\ee
The so-called ``Berry phase contribution'' to the path integral is obtained adding up this contribution over each site of the unit cell:
\be \label{berry}
	\fL_{\text{Berry}} = -\frac{1}{d}\sum_{j=1}^d \sum_{\alpha=1}^k p_\alpha \vec \phi^{\alpha,*}(j)\cdot \partial_\tau \vec{\phi}^\alpha(j).
\ee
Here $d$ is the size of the unit cell, and $\vec{\phi}^\alpha(j)$ is the field $\vec{\phi}^\alpha$ evaluate at site $j$. Since we are deriving a field theory about a classical ground state, to lowest order $\vec{\phi}^\alpha(j)$ is the colour of node $\alpha$ at site $j$. 

To obtain the complete quantum field theory, one must add to $\fL_{\text{Berry}}$ a gradient expansion of the SU($n$) lattice Hamiltonian, and this is where the lengthy calculations lie. However, if one is interested only in time-derivatives, it suffices to restrict attention to $\fL_{\text{Berry}}$, since the Hamiltonian is time independent. The lowest-order expansion of $\fL_{\text{Berry}}$, which amounts to replacing the $\vec{\phi}^\alpha(j)$ with their corresponding colour basis vectors $\vec{e}^\beta$ (where $\beta$ depends on $\alpha$ and $j$), will indicate how many of the sigma model's modes have linear dispersion. The next-order expansion, which takes into account the spatial fluctuations of the $\vec{\phi}^\alpha$ across the unit cell, will provide the topological angle content of the theory. 

\subsection{Dispersion Relations}

For each family of representations in Table~\ref{result}, we determine the lowest order contribution to $\fL_{\text{Berry}}$. For each field $\vec{\phi}^\alpha$ that is present, this indicates the presence of $(n-1)$ quadratically dispersing modes. Only in the case of a vanishing $\fL_{\text{Berry}}$ at this order does linear dispersion occur for each mode of the theory. However, even in this case, Lorentz invariance is not automatic since the fields $\vec{\phi}^\alpha$ will generically propagate with different velocities.

 Let us demonstrate how this works for the symmetric SU($n$) representations (corresponding to $k=1$). The ground states are very simple in this case: one row of $n$ nodes, with each colour occurring once. See for example (\ref{eq:sym1}). Therefore, we have 
\be  \label{eq:disperse}
	\fL_{\text{Berry}} = -
	\frac{p_1}{n}\sum_{j=1}^n \vec{\phi}^{\alpha,*}(j)\cdot\partial_\tau \vec{\phi}^\alpha(j)  \ho
\ee
where H.O. includes higher order terms. Since each colour occurs once in the sum on the RHS, the sum equals $\tr[U^\dag \partial_\tau U]$ for a unitary matrix $U$. In Appendix B of \cite{sun2019}, it is shown that this trace equals zero, so that all modes have linear dispersion in the representations with $k=1$.

For the remaining representations, we have compiled our results in 
Table~\ref{result2} and Table~\ref{result3}. Each row corresponds to a family of SU($n$) representations. The `Min \#' column counts the minimum number of $\mathbb{C}^n$ fields $\vec{\phi}^\alpha$ that have linear dispersion in the corresponding $\fm$ sigma model. The larger, right-hand column lists the conditions that the representation parameters $p_\alpha$ must satisfy in order for additional fields to acquire linear dispersion. Each condition is accompanied by a number in parenthesis: this dictates how many fields $\vec{\phi}_\alpha$ become linearly dispersing when this condition is satisfied. For example, the second row of Table~\ref{result2} corresponds to representations with $n-1$ rows in their Young tableaux. These representations will always have at least two linearly dispersing fields in their sigma model. In order to have more linearly dispersing fields, we must start to satisfy conditions. When $n$ is even, these conditions are $p_\alpha + p_{n-\alpha +1} = p_1$, for $\alpha=2,\cdots, \frac{n}{2}$. Each satisfied condition adds 2 more linearly dispersing fields to the sigma model. It is amusing to note that when all of these conditions are satisfied, we obtain the set of self-conjugate representations of SU($n$) (that don't have two rows of the same length).

For the detailed calculations that support the results in these tables, we refer the reader to \ref{app:disperse}. It is important to note that there is some ambiguity in the number of linear vs. quadratic modes, which follows from the trace identity $\tr[U^\dag\partial_\tau U]=0$. This expression allows us to rewrite a partial sum $\sum_{\alpha} \vec{\phi}^\alpha \cdot \partial_\tau \vec{\phi}^\alpha$ in terms of the the $\vec{\phi}^\beta$ that do not occur in the sum:
\be \label{eq:amb}
	\sum_{\alpha \in A} \vec{\phi}^\alpha \cdot \partial_\tau \vec{\phi}^\alpha =
	-\sum_{\beta \not\in A} \vec{\phi}^\beta \cdot \partial_\tau \vec{\phi}^\beta.
\ee
To be consistent, we will always choose to write the Berry phase contribution in terms of the least number of fields possible. However, of primary interest to us in this paper are theories that only have linearly-dispersing modes; in this case, the counting becomes uniquely defined. 

\begin{center}
\begin{table}[!h]
\centering
\begin{tabular}{| c | c | c  r |} \hline
Representation & Min \#  & Conditions & \\
\hline \hline 
$k=1$ & $n$   & none &  \\ \hline

\multirow{2}{*}{\vspace{-10mm} $k=n-1$} & \multirow{2}{*}{\vspace{-10mm} $2$} & 
$p_{\alpha} + p_{n-\alpha+1} = p_1 \hspace{5mm} (2) $& $n$ even; $\alpha=2,\cdots, \frac{n}{2}$ \\ \cline{3-4}
& & \parbox{1cm}{\begin{align*}\vspace{-3mm}
p_{\alpha} + p_{n-\alpha+1} &= p_1  \hspace{5mm} (2) \\
2p_{\frac{n+1}{2}} &= p_1 \hspace{5mm} (1) \\
\end{align*}} \vspace{-5mm}  & $n$ odd; $\alpha=2,\cdots, \frac{n-1}{2}$ \\
\hline

\multirow{2}{*}{\vspace{-10mm} $n=k\lambda$} & \multirow{2}{*}{\vspace{-10mm}$2\lambda $}  & $ p_{\alpha} + p_{k+1-\alpha} =p_1+p_k \hspace{5mm} (2\lambda)$ &  $k$ even; $\alpha=2,\cdots,\frac{k}{2}$ \\ \cline{3-4}
&  & \parbox{2cm}{\begin{align*}
p_{\alpha} + p_{k+1-\alpha} &= p_1 +p_k \hspace{5mm} (2\lambda) \\
2p_{\frac{k+1}{2}} &= p_1+p_k \hspace{5mm} (\lambda) \\
\end{align*}} \vspace{-5mm}  & $k$ odd; $\alpha=2,\cdots, \frac{k-1}{2}$\\ \hline
$n=2\lambda+1, k=2$ & $\lambda+1$ & $\lambda p_1 = (\lambda+1)p_2 \hspace{5mm} (\lambda)$ &  \\  \hline

\multirow{2}{*}{\vspace{-10mm} $n=k\lambda +1$} & \multirow{2}{*}{\vspace{-15mm}$2\lambda $}  & \parbox{2cm}{
\begin{align*} p_\alpha + p_{k+2-\alpha}&=p_2 + p_k \hspace{5mm} (2\lambda)\\
(\lambda+1)( p_2 + p_k)&=2\lambda p_1 \hspace{5mm} (\lambda+1)\\ 
\end{align*}}
 &  $k$ odd; $\alpha=3,\cdots,\frac{k+1}{2}$ \\ \cline{3-4}
$\lambda = \text{ even}, k>2$ &  & \parbox{2cm}{\begin{align*}
 p_{\alpha} + p_{k+2-\alpha} &=p_2+p_k \hspace{5mm} (2\lambda) \\
2p_{\frac{k+2}{2}} &= p_2+p_k \hspace{5mm} (\lambda ) \\
(\lambda+1)(p_2+p_k) &= 2\lambda p_1 \hspace{5mm} (\lambda+1) \\
\end{align*}} \vspace{-5mm}  &\hspace{15mm} $k$ even; $\alpha=3,\cdots, \frac{k}{2}$\\ \hline

\multirow{2}{*}{\vspace{-10mm} $n=k\lambda +1$} & \multirow{2}{*}{\vspace{-15mm}$2\lambda $}  & \parbox{2cm}{
\begin{align*} p_\alpha + p_{k+2-\alpha}&=p_2 + p_k \hspace{5mm} (2\lambda)\\
(\lambda+1)( p_2 + p_k)&=2\lambda p_1 \hspace{5mm} (\lambda+1)\\ 
\end{align*}}
 &  $k$ odd; $\alpha=3,\cdots,\frac{k+1}{2}$ \\ \cline{3-4}
$\lambda = \text{ odd}, k>2$ &  & \parbox{2cm}{\begin{align*}
 p_{\alpha} + p_{k+2-\alpha} &=p_2+p_k \hspace{5mm} (2\lambda) \\
2p_{\frac{k+2}{2}} &= p_2+p_k \hspace{5mm} (\lambda ) \\
(\lambda+1)(p_2+p_k) &= \lambda p_1 \hspace{5mm} (\lambda+1) \\
\end{align*}}  &\hspace{15mm} $k$ even; $\alpha=3,\cdots, \frac{k}{2}$\\ \hline \end{tabular}
\caption{Classification of dispersion relations in $\fm$ sigma models, Part I.}\label{result2}
\end{table}

\begin{table}[p]
\centering
\begin{tabular}{|c |c |c  r |}
\hline
Representation & Min \#  & Conditions &\\
\hline \hline 
\multirow{2}{*}{ \vspace{-15mm}\parbox{2cm}{\begin{align*} n &= k\lambda +c \\
c &= \text{ even} \\
c &\not = k-1 \\
k &>1 \\
\end{align*}}} & \multirow{2}{*}{\vspace{-20mm}$2(\lambda +1)$}  & \vspace{-3mm} \parbox{2cm}{\begin{align*}
p_\alpha + p_{c+1-\alpha} &= p_1+p_c \hspace{5mm} (2(\lambda+1))\\
(\lambda+1)(p_\beta+p_{k-\beta+c+1}) &= \lambda (p_1+p_c) \hspace{5mm} (2\lambda) \\
2(\lambda+1)p_{\frac{k+c+1}{2}} &= \lambda(p_1+p_c) \hspace{5mm} (\lambda) \\
\end{align*}} \vspace{-1mm}& \vspace{-3mm} \parbox{2cm}{\begin{align*} k \text{ odd} &\\ 
\alpha  =2,\cdots,\frac{c}{2}  &\\
\beta -c = 1,\cdots,  \frac{k-c-1}{2} &\\ \end{align*}} \vspace{-3mm} \\ \cline{3-4}
&  & \parbox{2cm}{\begin{align*}
p_\alpha + p_{c+1-\alpha} &= p_1+p_c \hspace{5mm} (2(\lambda+1))\\
(\lambda+1)(p_\beta+p_{k-\beta+c +1}) &= \lambda (p_1+p_c) \hspace{5mm} (2\lambda) \\
\end{align*}} \vspace{-3mm}& \parbox{2cm}{\begin{align*} \hspace{10mm}k  \text{ even}; 
\alpha =2,\cdots,\frac{c}{2}  &\\
\beta = c+1,\cdots, c+ \frac{k-c}{2} &\\ \end{align*}} \vspace{-3mm} \\ \hline

\multirow{2}{*}{ \vspace{-2mm}\parbox{2cm}{\begin{align*} n &= k\lambda +c \\
c &= \text{ odd} \\
c &\not = k-1 \\
k &>1 \\
\end{align*}}} & \multirow{2}{*}{\vspace{-15mm}$2(\lambda +1)$}  & \vspace{-2mm} \parbox{2cm}{\begin{align*}
p_\alpha + p_{c+1-\alpha} &= p_1+p_c \hspace{5mm} (2(\lambda+1))\\
2p_{\frac{c+1}{2}} &= p_1+p_c \hspace{5mm} (\lambda+1) \\
(\lambda+1)(p_\beta+p_{k-\beta+c+1}) &= \lambda (p_1+p_c) \hspace{5mm} (2\lambda) \\
2(\lambda+1)p_{\frac{k+c+1}{2}} &= \lambda(p_1+p_c) \hspace{5mm} (\lambda) \\
\end{align*}} \vspace{-1mm}& \vspace{-3mm} \parbox{2cm}{\begin{align*} k \text{ even} &\\ 
\alpha  =2,\cdots,\frac{c-1}{2}  &\\
\beta -c = 1,\cdots,  \frac{k-c-1}{2} &\\ \end{align*}} \vspace{-3mm} \\ \cline{3-4}
&  & \parbox{2cm}{\begin{align*}
p_\alpha + p_{c+1-\alpha} &= p_1+p_c \hspace{5mm} (2(\lambda+1))\\
2p_{\frac{c+1}{2}} &= p_1+p_c \hspace{5mm} (\lambda+1) \\
(\lambda+1)(p_\beta+p_{k-\beta+c +1}) &= \lambda (p_1+p_c) \hspace{5mm} (2\lambda) \\
\end{align*}} \vspace{-3mm}& \parbox{2cm}{\begin{align*} \hspace{10mm}k  \text{ odd}; 
\alpha =2,\cdots,\frac{c-1}{2}  &\\
\beta = c+1,\cdots, c+ \frac{k-c}{2} &\\ \end{align*}} \vspace{-5mm} \\ \hline

\multirow{2}{*}{\vspace{-15mm} $n=\lambda k + (k-1)$} & \multirow{2}{*}{\vspace{-10mm} $2(\lambda+1)$} &  \vspace{-2mm}  \parbox{2cm}{\begin{align*}
p_\alpha + p_{k-\alpha} &= p_1+p_{k-1} \hspace{5mm} (2(\lambda+1))\\
(\lambda+1)p_k &= \lambda(p_1+p_{k-1}) \hspace{5mm} (\lambda) \\
\end{align*} } \vspace{-5mm}  & $k \text{ odd}; \alpha=2,\cdots,\frac{k-1}{2} $
\\ \cline{3-4}
$\lambda \text{ even}$ & & \parbox{2cm}{\begin{align*}
p_\alpha + p_{k-\alpha} &= p_1+p_{k-1} \hspace{5mm} (2(\lambda+1))\\
(\lambda+1)p_k &= \lambda(p_1+p_{k-1}) \hspace{5mm} (\lambda) \\
2p_{\frac{k}{2}} &= p_1+p_{k-1} \hspace{5mm} (\lambda+1) \\
\end{align*} } \vspace{-5mm} &  $k \text{ even}; \alpha = 2, \cdots, \frac{k-2}{2}$ \\ \hline

\multirow{2}{*}{\vspace{-15mm} $n=\lambda k + (k-1)$} & \multirow{2}{*}{\vspace{-10mm} $2(\lambda+1)$} &  \vspace{-2mm}  \parbox{2cm}{\begin{align*}
p_\alpha + p_{k-\alpha} &= p_1+p_{k-1} \hspace{5mm} (2(\lambda+1))\\
(\lambda+1)p_k &= \lambda(p_1+p_{k-1}) \hspace{5mm} (\lambda) \\
\end{align*} } \vspace{-5mm}  & $k \text{ odd}; \alpha=2,\cdots,\frac{k-1}{2} $
\\ \cline{3-4}
$\lambda \text{ odd}$ & & \parbox{2cm}{\begin{align*}
p_\alpha + p_{k-\alpha} &= p_1+p_{k-1} \hspace{5mm} (2(\lambda+1))\\
2(\lambda+1)p_k &= \lambda(p_1+p_{k-1}) \hspace{5mm} (\lambda) \\
2p_{\frac{k}{2}} &= p_1+p_{k-1} \hspace{5mm} (\lambda+1) \\
\end{align*} } &  $k \text{ even}; \alpha = 2, \cdots, \frac{k-2}{2}$ \\ \hline
\end{tabular}
\caption{Classification of dispersion relations in $\fm$ sigma models, Part II.}
\label{result3}
\end{table}
\end{center}

\subsection{Topological Angles} \label{sub:top}

The next piece of information we can extract from the Berry phase contribution to the sigma model is the set of topological angles for each representation of SU($n$). This requires taking into account the spatial variation of the fields $\vec{\phi}^\alpha$ in each of the terms found in the previous section.

 As we have already seen, each field is associated with some condition on the Young tableaux parameters $p_\alpha$. When determining the set of topological angles, it will be important to keep track of these conditions; ultimately, this will lead to a list of angles for each of the conditions appearing in Table \ref{result2} and Table \ref{result3}. Our motivation for this bookkeeping will become apparent in the follow-up paper \cite{followup} when we introduce the flag manifold hierarchy that arises from mixed ferro- and antiferromagnetic order parameters: By `turning on' a subset of the $p_\alpha$ conditions, we will able to effectively reduce the symmetry of our sigma model from $\fm$ to $\SU(n')/[\tU(1)]^{n'-1}$, for some $n'<n$. It will be essential to keep track of which topological angles survive in the smaller theory.

To begin, we recall (\ref{berry}):
\be 
	\fL_{\text{Berry}} = -\frac{1}{d}\sum_{j=1}^d \sum_{\alpha=1}^k p_\alpha \vec \phi^{\alpha,*}(j)\cdot \partial_\tau \vec{\phi}^\alpha(j)
\ee
where $d$ is the unit-cell length. The spatially constant terms were analyzed in the previous section. Now, we take into account spatial fluctuations of the $\vec{\phi}^\alpha$. In this case, we may write the leading-order correction as
\be  \label{berry2}
	\fL_{\text{Berry}} = \cdots +  \epsilon_{\mu\nu} \frac{1}{d}\sum_{j=1}^d (j-1)\sum_{\alpha=1}^k p_\alpha \partial_\mu \vec \phi^{x(\alpha,j),*}\cdot \partial_\nu \vec{\phi}^{x(\alpha,j)}  \ho
\ee
where all of the terms are evaluated at the same lattice site, and $\cdots$ hides the terms from the previous section. The notation $x(\alpha,j)$ reminds us that for each field in the sum, we must consult the ground state structure (found in Section \ref{sec:ham}), and use both the row $(\alpha$) and column $(j$) to determine the index $x$. Using this, we may rewrite this contribution from the Berry phase term as 
\be
	\fL_{\text{Berry}} =\frac{1}{2\pi i} \sum_{\alpha=1}^n \theta_\alpha q_\alpha.
\ee
where 
\be
	q_\alpha := \epsilon_{\mu\nu}\partial_\mu\vec{\phi}^{\alpha,*}\cdot\partial_\nu \vec{\phi}^\alpha
\ee
is a total derivative. From here, we are able to read off the topological angles, $\theta_\alpha$. We will carry out this procedure for a few examples, and then refer the reader to \ref{app:angle} for the remaining calculations.

\begin{itemize}

\item Case 1: $k=1$ \newline In this case, the Berry phase term reduces to 
\be 
	\fL_{\text{Berry}} = \epsilon_{\mu\nu} \frac{p_1}{n}\sum_{j=1}^n (j-1) \partial_\mu \vec \phi^{\alpha,*}\cdot \partial_\nu \vec{\phi}^{\alpha}  
\ee
so that 
\be
	\theta_\alpha = \frac{2\pi p_1}{n}(\alpha-1).
\ee
Since there are no quadratically dispersing fields when $k=1$, these angles do not correspond to a nontrivial condition on the $p_\alpha$. 

\item Case 2: $k=n-1$\newline Starting from (\ref{berry2}), it is clear that we only have to focus on a single column in the coloured ground state diagram:
\be
	\fL_{\text{Berry}} = \cdots+ \frac{1}{2} \epsilon_{\mu\nu} \sum_{\alpha=1}^{n-1} p_\alpha \partial_\mu \vec{\phi}^{\alpha,*} \cdot\partial_\nu \vec{\phi}^{\alpha}
\ee
The topological angles are then
\be \label{eq:99ang}
	\theta_\alpha=\pi p_\alpha.
\ee
According to the conditions in Table~\ref{result2}, two fields are always linearly dispersing, corresponding to $\theta_1 =p_1\pi$ and $\theta_n = 0$. The remaining $n-2$ angles correspond to fields that must satisfy conditions on the $p_\alpha$. The exact relationship between angle and nontrivial condition is given below, making use of (\ref{eq:99ang}):

\begin{table}[h]
\centering
\begin{tabular}{|c |c | c c |} \hline
Subcase & Condition & Angles & \\ \hline \hline 
$n$ even & \parbox{1cm}{\begin{align*} p_\alpha + p_{n-\alpha+1}  = p_1\end{align*}} & $\theta_\alpha, \theta_{n-\alpha+1}$ & $\hspace{10mm}  \alpha=2,\cdots, \frac{n}{2}$ \\ \hline
$n$ odd &  \parbox{1.5cm}{\begin{align*} p_\alpha + p_{n-\alpha+1}  &= p_1 \\
2p_{\frac{n+1}{2}} &= p_1 \\
\end{align*}} & \parbox{1.5cm}{\begin{align*} \theta_\alpha,\theta_{n-\alpha+1} \\
\theta_{\frac{n+1}{2}} \\
\end{align*}} &
\hspace{10mm} $\alpha=2,\cdots,\frac{n-1}{2}$ \\ \hline
\end{tabular} 
\end{table}

\end{itemize}

In \ref{app:angle}, we carry out this procedure for the remaining representations of SU($n$). In Table~\ref{result4}, we collect those results and record all possible topological angles for each case. The relationships between angle and conditions on the $p_\alpha$ can be found in various tables in \ref{app:angle}.

\begin{table}[!h]
\centering 
\begin{tabular}{l | l c } 
Representation &  Topological Angles  \\ \hline \hline
$k=1$ & $\theta_\alpha = \frac{2\pi p_1}{n}(\alpha-1)$ \vspace{1mm} & $\alpha=1,2,\cdots, n$ \\ \hline
$k=n-1$ & $\theta_\alpha = \pi p_\alpha$ \vspace{1mm} & 
$\alpha=1,2,\cdots,n$ \\ \hline
$k=\frac{n}{\lambda}$ & $\theta_{\alpha,j} = \frac{\pi(p_\alpha + p_{k+1-\alpha})}{\lambda}(j-1) + \pi p_{k+1-\alpha}$ \vspace{1mm} &
$\alpha=1,\cdots, k$ \\ \hline
$n=2\lambda+1, k=2$ & \parbox{2cm}{\begin{align*} \theta_t &=  \frac{2\pi p_1}{\lambda+1}(t-1) + \pi p_1(\lambda-1) \\
\tilde \theta_j &= \frac{2\pi p_2}{\lambda} (j-1) + \pi p_2\lambda \\
\end{align*}} \vspace{-5mm} & \\ \hline
\parbox{2cm}{\begin{align*} n &= k\lambda+1, k>2 \\
\lambda &=\text{ even}, \lambda >1 \\
\end{align*}} &
\parbox{2cm}{\begin{align*}\theta_{\alpha,j} &= \frac{\pi(p_\alpha + p_{k+2-\alpha})}{\lambda}(j-1) +\pi p_\alpha  \\ \theta_t &= \frac{2\pi p_1}{\lambda+1}(t-1) + \pi p_1 \\
\end{align*}} \vspace{-5mm} & $\alpha=2,\cdots, k$ \\ \hline 
\parbox{2cm}{\begin{align*} n &= k\lambda+1, k>2 \\
\lambda &=\text{ odd}, \lambda >1 \\
\end{align*}} \vspace{-3mm} &
\parbox{2cm}{\begin{align*}\theta_{\alpha,j} &= \frac{\pi(p_\alpha + p_{k+2-\alpha})}{\lambda}(j-1) +\pi p_\alpha + \frac{\pi (p_\alpha +p_{k+2-\alpha})}{2}(\lambda-1) \\ \theta_t &= \frac{2\pi p_1}{\lambda+1}(t-1)  \\
\end{align*}} \vspace{-3mm} & $\alpha=2,\cdots, k$ \\ \hline 
$n=k\lambda+c, c\not=1, k-1 $ &
\parbox{2cm}{\begin{align*}
\theta_{\alpha,t} &= \frac{\pi(p_\alpha + p_{c-\alpha+1})}{\lambda+1}(t-1) + \pi\lambda p_{c-\alpha+1} + \pi(\lambda-1)p_\alpha  \\
\tilde \theta_{\beta,j} &= \frac{\pi (p_\beta+p_{k-\beta+c+1})}{\lambda}(j-1) + \pi(\lambda+1)p_{k-\beta+c+1} +\pi\lambda p_\beta \\
\end{align*}} \vspace{-3mm} & \parbox{2cm}{\begin{align*} \alpha &= 1,\cdots, c \\
\beta &= c+1,\cdots, k \\
\end{align*}} \vspace{-3mm}\\ \hline
\parbox{2cm}{\begin{align*} n &= k\lambda+(k-1) \\
\lambda &=\text{ odd} \\
\end{align*}} \vspace{-3mm} &
\parbox{2cm}{\begin{align*}
\theta_{\alpha,t} &= \frac{\pi(p_\alpha + p_{k-\alpha})}{\lambda+1}(t-1) +  \pi p_{k-\alpha} \\
\theta_j &= \frac{2\pi p_k}{\lambda}(j-1) + p_k\pi \\
\end{align*}} \vspace{-3mm} &
$\alpha =1,\cdots, k-1$ \\ \hline
\parbox{2cm}{\begin{align*} n &= k\lambda+(k-1) \\
\lambda &=\text{ even} \\
\end{align*}} &\parbox{2cm}{\begin{align*}
\theta_{\alpha,t} &= \frac{\pi(p_\alpha + p_{k-\alpha})}{\lambda+1}(t-1) +  \pi p_{k-\alpha} + \frac{\pi(p_\alpha+p_{k-\alpha})}{2}(\lambda-2) \\
\theta_j &= \frac{2\pi p_k}{\lambda}(j-1)  \\
\end{align*}} &
$\alpha =1,\cdots, k-1$ \\ \hline
\end{tabular}
\caption{Possible topological angles for various representations of SU($n$) chains. The index $j$ runs from 1 to $\lambda$ and the index $t$ runs from 1 to $\lambda+1$. These angles can often be simplified by using the freedom of shifting each angle by the same constant.}
\label{result4}
\end{table}

\section{A New Generalization of Haldane's Conjecture}
\label{sec:result}

The lengthy analysis of the previous section makes clear the fact that most representations of SU($n$)  chains do not lead to linearly-dispersing sigma models, as is the case for the symmetric SU($n$) chains \cite{LajkoNuclPhys2017, sun2019}. This already occurs in SU(3), for any representation that is neither self-conjugate nor completely symmetric: in this case, at least one (and at most two) of the $\CP^2$ fields $\vec
{\phi}^\alpha$ has quadratic dispersion. In order to achieve a purely linearly-dispersing theory, a series of constraints on the Young tableaux parameters $p_\alpha$ must be satisfied. In the special case of representations with all $p_\alpha$ nonzero and distinct, these constraints lead to the self-conjugate representations of SU($n$).

In a follow-up paper\cite{followup} we will consider in great detail these sigma models with both linear and quadratic dispersion relations. For now, we restrict to representations of SU($n$) that satisfy the various constraints listed in Table~\ref{result2} and Table~\ref{result3}.

Before proceeding further, we must reflect on what we are hoping to achieve with this classification. Ultimately, we are interested in the possible gapless phases in SU($n$) chains, and how one might extend Haldane's conjecture to novel representations. Based on our understanding of the symmetric models, we know that this task can be recast in terms of 't Hooft anomaly matching. The recipe is as follows:
\begin{itemize}
\item Step 1: Map an SU($n$) chain to a (linear-dispersing) flag manifold sigma model at low energies.
\item Step 2: Identify the 't Hooft anomalies of the sigma model. When such an anomaly is present, we may conclude that the ground state either exhibits spontaneously broken symmetry, or gapless excitations. 
\end{itemize}
In \cite{Tanizaki:2018xto} and \cite{ohmori2019sigma}, it was shown that an 't Hooft anomaly occurs in the $\fm$ sigma model when an additional $\mathbb{Z}_n$ discrete symmetry is present. This symmetry acts on the $n$ complex fields transitively according to 
\be \label{eq:action}
	\mathbb{Z}_n: \vec{\phi}^\alpha \mapsto \vec{\phi}^{\alpha+1}.
\ee
In \cite{meron}, this gapless property of SU($n$) chains with 't Hooft anomalies was reinterpreted in terms of fractional instantons. Indeed, it was shown that in the $\fm$ sigma model, topological excitations exist that generate a finite energy gap above the ground state, much in the same way that vortices drive the familiar Kosterlitz-Thouless transition.\cite{Kosterlitz_1973} For a certain set of topological angles $\{\theta^*_\alpha\}$, these excitations destructively interfere with each other and the mass-generating mechanism breaks down, thus leading to a gapless ground state. It turns out that when the 't Hooft anomaly is present, the topological angle content in the sigma model is precisely $\{\theta^*_\alpha\}$. Ultimately, this follows from the form of the action (\ref{eq:action}). Therefore, in addition to concerning ourselves with linear dispersion, we also restrict focus to SU($n$) representations whose translational symmetry group $\mathbb{Z}_d=\mathbb{Z}_n$, and acts transitively on the fields $\vec{\phi}^\alpha$.


Of course, it is important to acknowledge this is by no means an exhaustive classification of gapless phases in SU($n$) chains. We do not attempt to classify all possible 't Hooft anomalies in these models, and so we are limited to the current list of known anomalies, and apply this knowledge to our theories. Moreover, we must also remember that the absence of an anomaly teaches us nothing: we are unable to predict any ground state properties when this is the case. However, we do have the Lieb-Shultz-Mattis-Affleck (LSMA) theorem,\cite{lsm1,lsm2} which predicts either a gapless ground state or spontaneously broken symmetry for an SU($n$) chain whenever the sum $p := \sum_\alpha p_\alpha$ is not a multiple of $n$.

Having made these remarks, we are now in a position to seek out representations of SU($n$) that may be amenable to a generalized Haldane conjecture. We assume that all of the constraints on the Young tableaux parameters $p_\alpha$ have been satisfied, so that all of the $n$ fields $\vec{\phi}^\alpha$ are linearly dispersing. For each class of representation occurring in Table~\ref{result}, we record when it is possible for the translation group to equal $\mathbb{Z}_n$, and act transitively on the set of fields.
\begin{itemize}
\item Case 1: $k=1$. \newline Since $\mathbb{Z}_d=\mathbb{Z}_n$, this is possible for all $n$.
\item Case 2: $k=n-1$ \newline Since $\mathbb{Z}_d=\mathbb{Z}_2$, this is possible only in SU(2) (which reduces to Case 1).
\item Case 3: $k=\frac{n}{\lambda}, \lambda < n$ \newline Since $d= 2\lambda$, and $\lambda \leq \frac{n}{2}$, this is possible only when $k=2$. See the ground state in (\ref{eq:z4}) to understand how this comes about. In other words, when $n$ is even, Young tableaux with two rows (of differing lengths) give rise to flag manifold sigma models with an additional $\mathbb{Z}_n$ symmetry. According to Table~\ref{result2}, such representations are always linearly dispersing, so no other assumption on the row lengths $p_\alpha$ is required. Note that the angles in this case are (see Table~\ref{result4})
\be
	\theta_{\alpha}= \frac{2\pi}{n} (p_1+p_2) (\alpha-1) \hspace{10mm} \alpha=1,2,\cdots, n,
\ee

where we have shifted each angle by the constant $\pi p_1$.
\item Case 4: $n=2\lambda+1, k=2$ \newline Since $n$ cannot equal $d=\lambda(\lambda+1)$,  no such representations give rise to a $\mathbb{Z}_n$ symmetry. 

\item Case 5: $ n =k\lambda +1, k>2, \lambda>1$. \newline
In this case, the $\mathbb{Z}_d$ symmetry does not act transitively on the set of $n$ fields: Some of the fields lie in an orbit of order $\lambda+1$, while the remaining fields lie in orbits of size $2\lambda$.

\item Case 6$: n=k\lambda + c, c\not=1,k-1$. \newline
Similar to case 5, the fields do not lie in a single orbit under the action of $\mathbb{Z}_d$. So while it is possible for $\mathbb{Z}_d=\mathbb{Z}_n$, the fields do not transform under the necessary action (\ref{eq:action}). The simplest example of this is SU(12) with $k=5$ rows in a Young diagram. Under the $\mathbb{Z}_n$ action, the fields partition into three orbits of size 6,4 and 2, and the current anomaly classification is no longer applicable. 

\item Case 7$: n=k\lambda + (k-1)$. \newline
Similar to the previous two cases: the fields do not lie in a single orbit under the action of $\mathbb{Z}_d$.

\end{itemize} 

In summary, we find only one new family of SU($n$) representations that give rise to a linearly-dispersing $\fm$ flag manifold sigma model with the $\mathbb{Z}_n$ symmetry (\ref{eq:action}). It is the set of representations with two rows (of different lengths) in their Young tableaux, when $n$ is even. The corresponding topological angles in this theory are 
\be
	\theta_\alpha =\frac{2\pi}{n} (p_1+p_2)\alpha \hspace{10mm} \alpha=1,2,\cdots, n
\ee
so that $p_1+p_2$ plays the role of $p_1$ in the symmetric models. According to the results in \cite{sun2019}, we may conclude that an 't Hooft anomaly is present whenever $p_1+p_2$ is not a multiple of $n$. This is also consistent with the LSMA theorem, mentioned above.\cite{lsm1,lsm2} Moreover, based on the classification of SU($n$) WZW flows in \cite{Yao:2018kel}, we may further conclude that only when $(p_1+p_2)$ is coprime with $n$ is a stable gapless phase possible. Otherwise, if $p_1+p_2$ shares a nontrivial common divisor with $n$, then the theory is necessarily gapped with spontaneously broken symmetry. 

On the other hand, when $p_1+p_2$ is a multiple of $n$ (and the LSMA theorem does not apply), it should be possible to have a unique, translationally invariant ground state with a finite energy gap. This statement is supported by the fact that when $p_1+p_2=n$, it is straightforward to write down a translationally invariant AKLT state.\cite{lsm2, AKLT1988} Indeed, we may construct a singlet over $p_1$ consecutive sites using $(p_1-p_2)$ fundamentals and $p_2$ antisymmetric doublets. Since each site has $p_1$ representations (either fundamentals or doublets), we may shift the singlet by one site as we move down the rows of the valence bond solid. 

As an example, let us explain this construction in greater detail for the case of SU(4), with $p_1=3$ and $p_2=1$. We denote by $\alpha_i^a$ a fundamental representation of SU(4) at site $i$. Then on each site of the chain, we have the representation 
\be
	|\alpha^1_i,\alpha^2_i,\alpha^3_i;\alpha^4_i\rangle,
\ee
which is symmetric under permutations of the first three entries, and antisymmetric under exchanges with the fourth entry. For instance, 
\[
	|\alpha^1_i,\alpha^2_i,\alpha^3_i;\alpha^4_i\rangle = |\alpha^2_i,\alpha^1_i,\alpha^3_i;\alpha^4_i\rangle
	=|\alpha^2_i,\alpha^3_i,\alpha^1_i;\alpha^4_i\rangle,
\]
\[
	|\alpha^1_i,\alpha^2_i,\alpha^4_i;\alpha^3_i\rangle = -|\alpha
	^1_i,\alpha^2_i,\alpha^3_i;\alpha^4_i\rangle = -|\alpha^1_i,\alpha^4_i,\alpha^3_i;\alpha^2_i\rangle
\]
\begin{figure}[h]
\centering
\includegraphics[width=.7\textwidth]{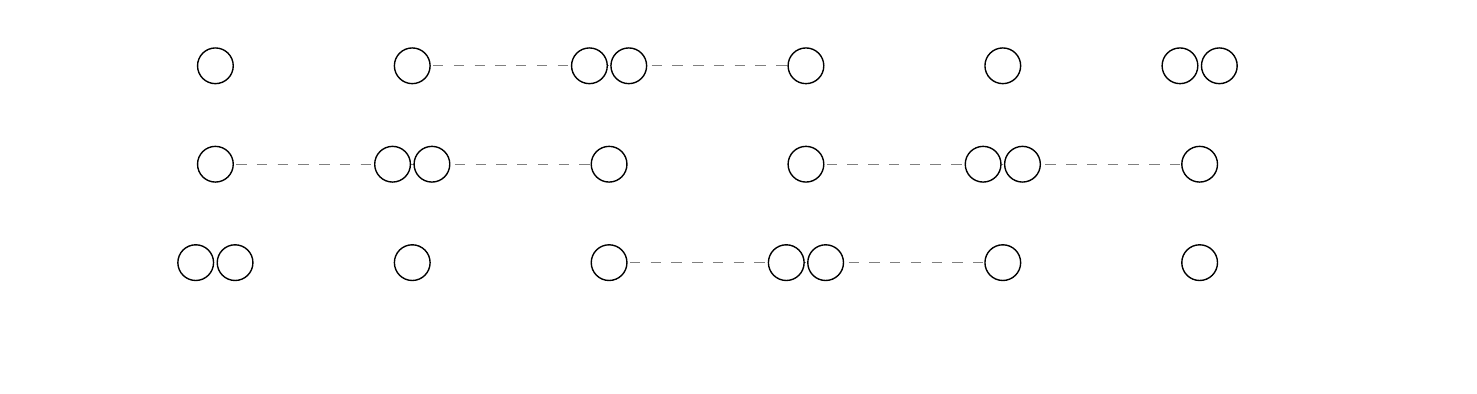}
\caption{AKLT state for an SU(4) chain with $p_1=3$ and $p_2=1$. Singlets are formed from three consecutive sites, using two fundamentals (single circles), and one antisymmetric doublet (double circle). }
\label{fig:aklt1}
\end{figure}
Using two fundamental representations $\alpha^a_i$, and one antisymmetric doublet representation $|\alpha^a_i;\alpha^b_i \rangle = -|\alpha^b_i ;\alpha^a_i\rangle$, we may contract indices to form a singlet across three sites according to 
\be
	\epsilon_{\alpha^1_i \alpha^2_{i+1} \alpha^3_{i+1} \alpha^4_{i+2}} |\alpha^a_i, \alpha^1_i, \alpha^b_i; \alpha^c_i\rangle |\alpha^d_{i+1},\alpha^{e}_{i+1},\alpha^2_{i+1};\alpha^3_{i+1}\rangle |\alpha^4_{i+2},\alpha^f_{i+2},\alpha^g_{i+2};\alpha^h_{i+2}\rangle.
\ee
Here $\epsilon_{1234}$ is the antisymmetric tensor. The remaining free representations $\alpha^a_i,\cdots,\alpha^h_{i+2}$  are then contracted into different singlet bonds, over different sets of three sites. By using the pattern shown in Fig~\ref{fig:aklt1},  a translationally invariant valence bond solid can be constructed, that is also parity symmetric. In fact, for general even $n$, we may always choose a $p_1$-site singlet bond that is symmetric under parity, leading to a parity-symmetric AKLT state. See Fig~\ref{fig:aklt2} and Fig~\ref{fig:aklt3} for two additional examples in SU(6).

\begin{figure}[h]
\centering
\includegraphics[width=.9\textwidth]{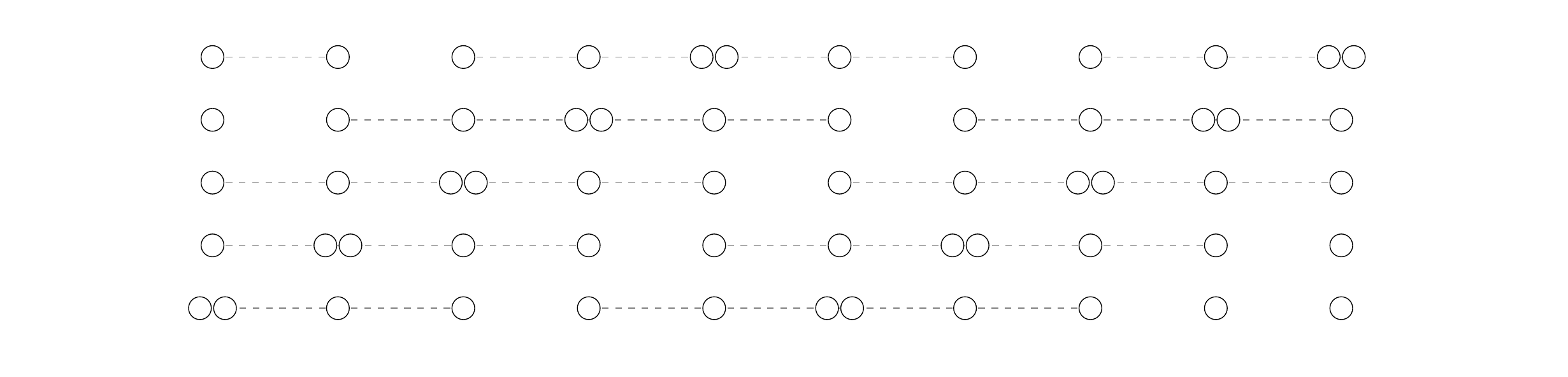}
\caption{AKLT state for an SU(6) chain with $p_1=5$ and $p_2=1$. Singlets are formed from five consecutive sites, using four fundamentals (single circles), and one antisymmetric doublet (double circle). }
\label{fig:aklt2}
\end{figure}

\begin{figure}[h]
\centering
\includegraphics[width=.9\textwidth]{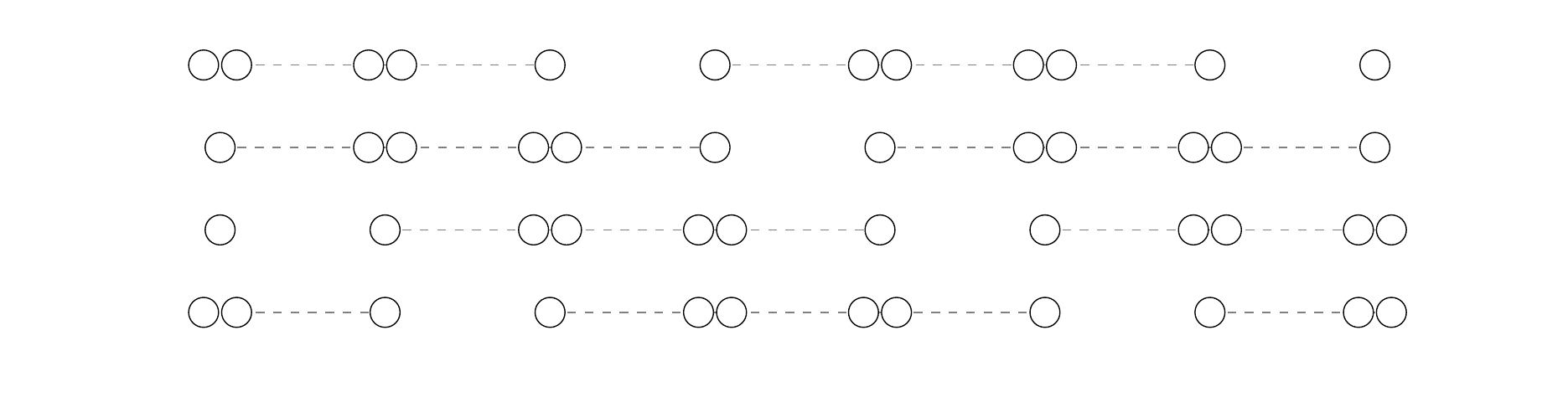}
\caption{AKLT state for an SU(6) chain with $p_1=4$ and $p_2=2$. Singlets are formed from four consecutive sites, using two fundamentals (single circles), and two antisymmetric doublet (double circle). }
\label{fig:aklt3}
\end{figure}

\section{Conclusions}

In this work, we have attempted to classify all SU($n$) chains that may admit a mapping to the $\fm$ flag manifold sigma model at low energies. Unless two rows of the Young tableaux have the same length, it seems possible to realize such a mapping for any irreducible representation. However, in most cases the corresponding sigma model will have complex fields $\vec{\phi}^\alpha \in \mathbb{C}^n$ with both quadratic and linear dispersion relations. One consequence of this is that Lorentz invariance can never emerge at low energies, as it does for the symmetric SU($n$) chains that posses only linearly dispersing fields. In a follow-up paper \cite{followup}, we will introduce a mechanism around this, which provides a new path to Lorentz invariance and also reveals a hierarchy of flag manifold sigma models for each value of $n$. For now, we have classified which representations lead to only linearly dispersing models, and have determined the topological angles in each case. Moreover, within this subset of representations, we have further classified which chains also admit a $\mathbb{Z}_n$ symmetry that acts transitively on the SU($n$) fields. This property is of interest, as it leads to the presence of an 't Hooft anomaly, and the possibility of generalizing Haldane's conjecture to new representations of SU($n$). In the end, we have found that only the SU($n$) irreps with even $n$ and two rows in their Young tableaux, with lengths $p_1\not=p_2$, satisfy all of these properties (in addition to the symmetric irreps considered previously). As a result, we have made the following modest extension of the SU($n$) generalization of Haldane's conjecture for even $n$: when $p_1+p_2$ is coprime with $n$, a gapless ground state is predicted; otherwise, a gapped ground state is expected, with spontaneously broken symmetry if $p_1+p_2$ is not a multiple of $n$.

\vspace{5mm}

\section*{Acknowledgements}

KW is supported by an NSERC PGS-D Scholarship, as well as by the Stewart Blusson Quantum Matter Institute's QuEST Program. IA is supported by NSERC of Canada Discovery Grant 04033-2016.

\vspace{5mm}

\noindent\textbf{References}
\vspace{3mm}

\bibliography{flag.bib}
\bibliographystyle{apsrev4-1}

\appendix

\section{Ground State Calculations} \label{app:order}

In this appendix, we prove that the coloured diagrams presented in Sec~\ref{sub:case2} do indeed minimize the classical Hamiltonian. The proofs for the remaining subsections of Sec~\ref{sec:ham} are similar, and we omit them here. 

Our task is to minimize 
\be
	\tr[ S(i)S(i+1)] = \sum_{\alpha,\beta=1}^{n-1} p_\alpha p_\beta |\vec{\phi}^{\alpha,*}(i)\cdot\vec{\phi}^\beta(i+1)|^2.
\ee
Using the orthonormal basis $\vec{e}^\alpha$, this expression reduces to 
\be \label{eq:min0}
	\tr[ S(i)S(i+1)] = \sum_{\alpha,\beta=1}^n p_\alpha p_\beta |\phi^\beta_\alpha(i+1)|,
\ee
where we've defined $p_n :=0$. Since $\vec{\phi}^\beta(i+1) = \vec{e}^{\alpha'}$ for some $\alpha'$ (and all of the $\alpha'$ are distinct), we may introduce a permutation operator on the set of $n$ elements, $\sigma: \{1,2,\cdots, n\} \to \{1,2,\cdots, n\}$ that obeys
\[
	 \vec{\phi}^\beta(i+1) = \vec{e}^{\sigma(\beta)}
\]
and rewrite (\ref{eq:min0}) as 
\be \label{eq:min}
	\tr[S(i)S(i+1)] =  \sum_{\alpha=1}^{n} p_\alpha p_{\sigma(\alpha)}.
\ee
Thus, our task amounts to finding the permutation $\sigma$ that minimizes (\ref{eq:min}). By defining a vector $\vec{x} := (p_1,p_2,\cdots, p_n)$, we can think of $\sigma$ as specifying a second vector $\vec{y}$; (\ref{eq:min}) is then their dot product. Since the entries of both $\vec{x}$ and $\vec{y}$ are nonnegative and nondegenerate, it is clear how to choose $\sigma$ so that $\vec{y}$ is as orthogonal to $\vec{x}$ as possible: 
\begin{itemize}
\item Since $p_1$ is the largest component of $\vec{x}$, we assign to $\sigma(1)$ the smallest possible component of $\vec{y}$, which is $p_n$.
\item Next, assign to $\sigma(2)$ the second smallest possible component of $\vec{y}$, which is $p_{n-1}$.
\item Repeating this procedure, we see that indeed the classical Hamiltonian is minimized by the reverse-ordered ground state, corresponding to a permutation operator
\[
	\sigma: i \mapsto k+1-i
\]
The basis states at site $i+1$ are $\vec{\phi}^\beta(i+1) = \vec{e}^{\sigma(\beta)}$. $\square$
\end{itemize}

\section{Dispersion Relation Calculations} \label{app:disperse}

In this appendix, we derive the results found in Tables~\ref{result2} and \ref{result3}. In each expression for $\fL_{\text{Berry}}$, we refrain from writing ``+ higher order terms" each time. The symmetric representations were already considered in the main text, so we begin with the $k=n-1$ representations. Systematically, we will consider all representations that appear in Table~\ref{result}, and record our results in Table~\ref{result2} and Table~\ref{result3}.

\begin{itemize}

\item Case 1: $k=n-1$ \newline
According to the pattern of ground states (see (\ref{eq:su51}) for example), two of the colours occur once (in the first position of the column), and the remaining $n-2$ colours occur twice, with reverse ordering. Therefore we have, according to (\ref{eq:disperse}),
\be
	\fL_{\text{Berry}} = -\frac{p_1}{2}(\vec{\phi}^{1,*}\cdot\partial_\tau \vec{\phi}^1 + \vec{\phi}^{2,*}\cdot\partial_\tau \vec{\phi}^2)
	-\frac{1}{2}
	\sum_{\alpha=3}^{n} (p_{\alpha-1} +p_{n-\alpha+2}) \vec{\phi}^{\alpha,*}(j) \cdot\partial_\tau \vec{\phi}^\alpha(j).
\ee
We have chosen to keep the $\vec{\phi}^\alpha$ notation, instead of the colour basis $\vec{e}^\alpha$, to remind ourselves that we are deriving terms in the field theory of fluctuations about the ground state. Using $\tr[U^\dag \partial_\tau U]=0$, this can be rewritten as 
\be
	\fL_{\text{Berry}} = 
	-\frac{1}{2}\sum_{\alpha=3}^{n} (p_{\alpha-1} +p_{n-\alpha+2} - p_1) \vec{\phi}^{\alpha,*}(j) \cdot\partial_\tau \vec{\phi}^\alpha(j).
\ee
Now we have up to $(n-2)$ fields with quadratic dispersion. The exact number will depend on how many of the conditions $p_{\alpha-1}+p_{n-\alpha+2}-p_1=0$ are satisfied. Each constraint corresponds to two fields $\vec{\phi}_\alpha$, except for the constraint $2p_{(n+1)/2}=p_1$ when $n$ is odd, which corresponds to a single field. The representations that satisfy every constraint, and thus give rise to sigma models with purely linear dispersion, correspond to the so-called self-conjugate representations of SU($n$) (that don't have equal row lengths in their tableaux). Indeed, in SU(3), the condition is $2p_2=p_1$, corresponding to Young diagrams which were previously considered in \cite{Wamer2019} (see Fig~\ref{yngew}). 
\begin{figure}[!h]
\centering
\includegraphics[width=.9\textwidth]{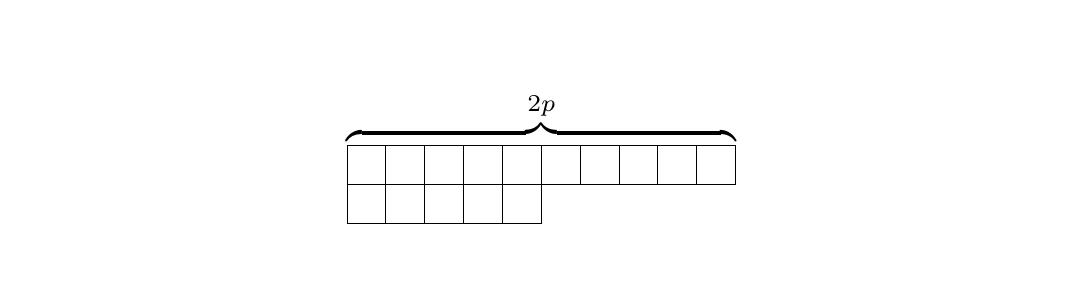}
\caption{Young diagram of self-conjugate representations in SU(3).}
\label{yngew}
\end{figure}
Similarly, in SU(4), the condition for linear dispersion is $p_2+p_3=p_1$, which is equivalent to the self-conjugate condition $p_1-p_2 =p_3$.

\item Case 2: $k=\frac{n}{\lambda}$

In this case, the ground states have order of length $2\lambda$, with each colour occurring twice. Let us adopt the notation
\be
	A^\alpha :=\vec{\phi}^{\alpha,*} \cdot \partial_\tau \vec{\phi}^\alpha.
\ee
Following the patterns (\ref{eq:nk1}) and (\ref{eq:nk2}) as a guide, we find that
\be
	\fL_{\text{Berry}} = -\frac{1}{2\lambda} \sum_{\alpha=1}^k \sum_{j=1}^\lambda (p_\alpha + p_{k+1-\alpha})A^{\alpha+(j-1)k)},
\ee
which can be rewritten using $\tr[U^\dag \partial_\tau U]=0$ to yield 
\be
	\fL_{\text{Berry}} = -\frac{1}{2\lambda} \sum_{\alpha=2}^{k-1} \sum_{j=1}^\lambda (p_\alpha + p_{k+1-\alpha} - p_1 - p_k)A^{\alpha+(j-1)k}.
\ee
This suggests that there can be up to $(k-2)\lambda$ fields with quadratic dispersion. In order to remove all of these modes, the representation must satisfy certain constraints. When $k$ is even, they are 
\be \label{txt1}
	p_\alpha + p_{k+1-\alpha} = p_1+p_k \hspace{10mm} \alpha=2,\cdots,\frac{k}{2}
\ee
and when $k$ is odd, they are
\begin{align} \label{txt2}
	p_\alpha + p_{k+1-\alpha} &= p_1+p_k \hspace{10mm} \alpha=2,\cdots,\frac{k-1}{2} \\
	2p_{\frac{k+1}{2}} &= p_1 + p_k \label{txt3}
\end{align}

Each of the constraints (\ref{txt1}), (\ref{txt2}) corresponds to the dispersion of $2\lambda$ fields, and the constraint (\ref{txt3}) corresponds to dispersion relation of $\lambda$ fields. In the special case of $k=2$, which corresponds to Young tableaux with two rows, we have automatically that $\fL_{\text{Berry}}=0$ for all values of $p_1$ and $p_2$. The simplest representation for larger $k$ is shown in Fig~\ref{su8}, for SU(8).
\begin{figure}[!ht]
\centering
\includegraphics[width=.9\textwidth]{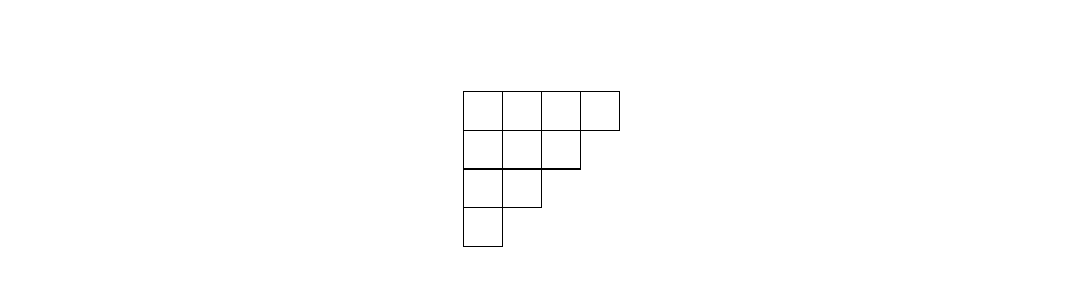}
\caption{This diagram satisfies the linear dispersion constraint $p_1+p_4=p_2+p_3$.}
\label{su8}
\end{figure}

\item Case 3: $n=2\lambda+1, k=2$ \newline

In this case the ground states have unit-cells of length $\lambda(\lambda+1)$. For example, in SU(5) with $k=2$, a candidate ground state is

\be
	\cir{e1} \hspace{5mm} \cir{e2} \hspace{5mm} \cir{e5} \hspace{5mm}  
	\cir{e1} \hspace{5mm} \cir{e2} \hspace{5mm} \cir{e5} \hspace{5mm} 
\ee
\vspace{-6mm}
\[
	\cir{e3} \hspace{5mm} \cir{e4} \hspace{5mm} \cir{e3} \hspace{5mm}  
	\cir{e4} \hspace{5mm}  \cir{e3} \hspace{5mm}  \cir{e4} \hspace{5mm} 
\]
For all of these representations, the first row will have periodicity $\lambda+1$, and the second row will have periodicity $\lambda$. The Berry phase term is then
\be
	\fL_{\text{Berry}} = -\frac{p_1}{(\lambda+1)}\sum_{\alpha=1}^{\lambda +1}A^\alpha
	-\frac{p_2}{\lambda} \sum_{\beta=1}^\lambda A^\beta.
\ee
which can be rewritten to give 
\be
	\fL_{\text{Berry}} = -\frac{1}{\lambda(\lambda+1)}\left[ p_2(\lambda+1) - p_1\lambda\right]	 \sum_{\beta=1}^\lambda A^\beta.
\ee
For most values of $p_1$ and $p_2$, the corresponding sigma model will have $\lambda$ fields with quadratic dispersion, and $\lambda+1$ fields with linear dispersion.\begin{footnote}{As mentioned below (\ref{eq:amb}), we can rewrite $\mathcal{L}_{\text{Berry}}$ to have $\lambda+1$ fields with quadratic dispersion, and $\lambda$ fields with linear dispersion.}\end{footnote} However, for the special representation satisfying
\be
	\lambda p_1 = (\lambda+1)p_2
\ee
a theory with purely linearly dispersing modes is achieved.

\item Case 4: $n=k\lambda+1$, with  $k>2$ and $\lambda>1 $ \newline
According to Table~\ref{result}, we must further specify the parity of $\lambda$:
\begin{itemize}
\item[\textbullet] $\lambda = \text{ even } $ \newline
The ground state unit-cell has length $2\lambda(\lambda+1)$ in this case. The first row has $(\lambda+1)$-site order, while the remaining $k-1$ rows have $2\lambda$-site order, with the coloured nodes exhibiting reverse ordering. Therefore, the Berry phase term is 
\be \label{eq:5a}
	\fL_{\text{Berry}} = -\frac{ p_1 }{(\lambda+1)}\sum_{\alpha=1}^{\lambda+1}A^\alpha
	 - \frac{1}{2\lambda} \sum_{\beta=2}^{k}\sum_{j=1}^\lambda(p_\beta + p_{k+2-\beta})A^{\beta + j(k-1)}.
\ee
While simpler to rewrite in terms of the $A^\alpha$ with $\alpha> \lambda+1$, we will always choose to rewrite the Berry phase in terms of the least number of fields possible. This leads to 
\be \label{eq:5a}
	\fL_{\text{Berry}} = -\frac{ 1 }{2\lambda (\lambda+1)}\sum_{\alpha=1}^{\lambda+1}\left[ 2\lambda p_1 - (\lambda+1)(p_2+p_k)\right] A^\alpha
	 - \frac{1}{2\lambda} \sum_{\beta=3}^{k-1}\sum_{j=1}^\lambda(p_\beta + p_{k+2-\beta} - p_2 - p_k)A^{\beta + j(k-1)} .
\ee

\item[\textbullet] $\lambda = \text{ odd } $ \newline
In this case, the result found in (\ref{eq:5a}) is slightly modified to 
\be \label{eq:5a}
	\fL_{\text{Berry}} = -\frac{ 1 }{\lambda (\lambda+1)}\sum_{\alpha=1}^{\lambda+1}\left[ \lambda p_1 - (\lambda+1)(p_2+p_k)\right] A^\alpha
	 - \frac{1}{\lambda} \sum_{\beta=3}^{k-1}\sum_{j=1}^\lambda(p_\beta + p_{k+2-\beta} - p_2 - p_k)A^{\beta + j(k-1)}.
\ee
\end{itemize}

\item Case 5: $n=k\lambda +c$ with $c\not= 1,k-1$ \newline
The ground states in the case have unit-cell order of length $2\lambda(\lambda+1)$. The Berry phase contribution is 
\be
	\fL_{\text{Berry}} = -\frac{1}{2(\lambda+1)}\sum_{\alpha=1}^c \sum_{j=1}^{\lambda+1}\left[ p_\alpha + p_{c+1-\alpha}\right]A^{\alpha + (j-1)c}
	 -\frac{1}{2\lambda }\sum_{\beta=c+1}^k \sum_{j=1}^{\lambda}\left[ p_\beta + p_{k-\beta+c+1}\right]A^{c(\lambda+1 ) + (\beta-c) + (j-1)(k-c)}.
\ee
Using the $\tr[U^\dag\partial U]=0$ identity, we can only remove $2(\lambda+1)$ fields. We are left with:
\be
	\fL_{\text{Berry}} = -\frac{1}{2(\lambda+1)}\sum_{\alpha=2}^{c-1}\sum_{j=1}^{\lambda+1}\left[ p_\alpha + p_{c+1-\alpha} - (p_1 + p_c) \right]A^{\alpha + (j-1)c}
\ee
\[
	 -\frac{1}{2\lambda(\lambda+1) }\sum_{\beta=c+1}^k \sum_{j=1}^{\lambda}\Big[(\lambda+1)( p_\beta + p_{k-\beta+c+1}) - \lambda (p_1+p_c)\Big]A^{c(\lambda+1 ) + (\beta-c) + (j-1)(k-c)}.
\]

The number of independent constraints that must be satisfied in order to achieve a sigma model with purely linear dispersion is $\lfloor \frac{c}{2}\rfloor$ + $\lfloor \frac{k-c}{2}\rfloor$.

\item Case 6: $n=k\lambda + (k-1)$ \newline

In this final case, we must again split into two cases, based on the parity of $\lambda$:
\begin{itemize}
\item[\textbullet] $\lambda = \text{ even } $ \newline
According to Table~\ref{result}, the ground state has $\lambda(\lambda+1)$-site order. The first $c-1$ rows of the ground state has period $2(\lambda+1)$, while the last row has period $\lambda$. The Berry phase term is 
\be \label{eq:6a}
	\fL_{\text{Berry}} = -\frac{1}{(\lambda+1)} \sum_{\alpha=1}^{k-1} \sum_{j=1}^{\lambda+1}[ p_\alpha + p_{k-\alpha}]A^{\alpha + (j-1)(k-1)}
	-\frac{1}{\lambda}\sum_{\beta=1}^{\lambda} p_k A^{(k-1)(\lambda+1) + \beta}
\ee
\[
	 = -\frac{1}{(\lambda+1)} \sum_{\alpha=2}^{k-2} \sum_{j=1}^{\lambda+1}[ p_\alpha + p_{k-\alpha} - p_1 -p_{k-1}]A^{\alpha + (j-1)(k-1)}
\]
\[
	-\frac{1}{\lambda(\lambda+1)}\sum_{\beta=1}^{\lambda}[ (\lambda+1)p_k  - \lambda(p_1+p_{k-1})]A^{(k-1)(\lambda+1) + \beta}.
\]

\item[\textbullet] $\lambda = \text{ odd } $ \newline
Now the ground state has $2\lambda(\lambda+1)$ order, which changes the result in (\ref{eq:6a}) to 
\be
	 = -\frac{1}{2(\lambda+1)} \sum_{\alpha=2}^{k-2} \sum_{j=1}^{\lambda+1}[ p_\alpha + p_{k-\alpha} - p_1 -p_{k-1}]A^{\alpha + (j-1)(k-1)}
\ee
\[
	-\frac{1}{2\lambda(\lambda+1)}\sum_{\beta=1}^{\lambda}[ 2(\lambda+1)p_k  - \lambda(p_1+p_{k-1})]A^{(k-1)(\lambda+1) + \beta}.
\]
\end{itemize}

\end{itemize}




\section{Topological Angle Calculations} \label{app:angle}

In this appendix, we continue the topological angle calculations that were started in Sec~\ref{sub:top}.

\begin{itemize}

\item Case 3: $k=\frac{n}{\lambda}$ \newline In this case, the correction to $\fL_{\text{Berry}}$ is
\be
	\fL_{\text{Berry}} = \frac{1}{2\lambda} \epsilon_{\mu\nu} \sum_{\alpha=1}^k \sum_{j=1}^\lambda \Big[ (j-1) p_\alpha + (\lambda+j-1) p_{k+1-\alpha}\Big]\partial_\mu \vec{\phi}^{\alpha,j,*}\cdot\partial_\nu \vec{\phi}^{\alpha,j}
\ee
so that the topological angles are
\be
	\theta_{\alpha,j} 
	= \frac{\pi}{\lambda} (p_\alpha + p_{k+1-\alpha})(j-1) + \pi p_{k+1-\alpha}
	\hspace{10mm}
	\alpha=1,\cdots, k; j=1,\cdots, \lambda
\ee
Here we use two indices to enumerate the fields. Since $2\lambda$ of the fields always have linear dispersion, the angles $\theta_{1,j}$ and $\theta_{k,j}$ are always present. Meanwhile, the remaining $(k-2)\lambda$ angles are associated to certain conditions, according to the following table:
\begin{table}[!h]
\centering
\begin{tabular}{|c |c |c c |} \hline
Subcase & Condition & Angles & \\ \hline \hline
$k$ even & \parbox{1cm}{\begin{align*} p_\alpha + p_{k+1-\alpha}  = p_1 + p_k \end{align*} }& $\theta_{\alpha,j}, \theta_{k-\alpha+1,j}$ & \hspace{10mm} $ \alpha=2,\cdots, \frac{k}{2}; j=1,\cdots,\lambda$  \\ \hline
$k$ odd & \parbox{2cm}{\begin{align*}
 p_\alpha + p_{k-\alpha+1}  & = p_1+p_k \\
 2p_{\frac{k+1}{2}} & = p_1 + p_k 
 \end{align*}}
 &
\parbox{2cm}{\begin{align*}
\theta_{\alpha,j}, \theta_{k-\alpha+1,j} \\
\theta_{\frac{k+1}{2},j} 
\end{align*} } & \hspace{10mm}
$ \alpha=2,\cdots, \frac{k-1}{2}, j=1,\cdots, \lambda$ \\ \hline
\end{tabular} 
\end{table}

\item Case 4: $n=2\lambda+1$ with $k=2$\newline
We again refer to Table~(\ref{result2}). From
\be
	\fL_{\text{Berry}} = \epsilon_{\mu\nu} \frac{p_1}{\lambda(\lambda+1)} \sum_{j=1}^{\lambda} \sum_{t=1}^{\lambda+1} (t-1 + (j-1)(\lambda+1))\partial_\mu \vec{\phi}^{t,*}\cdot\partial_\nu\vec{\phi}^t
\ee
\[
	+ \epsilon_{\mu\nu} \frac{p_2}{\lambda(\lambda+1)} \sum_{j=1}^{\lambda} \sum_{t=1}^{\lambda+1} (j-1 + (t-1)\lambda)\partial_\mu \vec{\phi}^{j+\lambda+1,*}\cdot\partial_\nu\vec{\phi}^{j+\lambda+1},
\]
we see that there are two families of angles:
\be
	\theta_t =\frac{2\pi p_1}{(\lambda+1)}(t-1) 
	+\pi p_1 (\lambda-1) \hspace{10mm} t=1,\cdots,\lambda+1
\ee
and
\be
	\tilde \theta_j = 
	\frac{2\pi p_2}{\lambda}(j  -1) +\pi p_2 \lambda
  \hspace{10mm} j =1,\cdots, \lambda
\ee

The $\theta_t$ angles correspond to fields that are always linearly dispersing. The remaining angles are associated to the single condition $\lambda p_1 = (\lambda+1)p_2$.

\item Case 5a: $n=k\lambda +1$  with $k>2$ and $\lambda$ even. \newline
Since the first row of the ground state has $(\lambda+1)$-site order, and the unit cell order is $2\lambda(\lambda+1)$, we may write as two parts $\fL_{\text{Berry}} = \fL_{\text{Berry}}^1 + \fL_{\text{Berry}}^2$. The first part is
\be
	\fL_{\text{Berry}}^1 = \epsilon_{\mu\nu} \frac{p_1}{2\lambda(\lambda+1)} \sum_{t=1}^{\lambda+1} \sum_{j=1}^{2\lambda}(t-1 +  (j-1)(\lambda+1))\partial_\mu\vec{\phi}^{t,*} \cdot\partial_\nu\vec{\phi}^t.
\ee
which gives rise to the following topological angles:
\be
	\theta_t = \frac{\pi p_1}{\lambda(\lambda+1)} \sum_{j=1}^{2\lambda} (t-1 + (j-1)(\lambda+1)) \hspace{10mm}
	t=1,2,\cdots,\lambda+1.
\ee
These angles can be further simplified to
\be
	\theta_t = \frac{2\pi p_1}{(\lambda+1)}(t-1)
	+ \pi p_1  
 \hspace{10mm}
	t=1,2,\cdots,\lambda+1.
\ee

The second part of the Berry phase corresponds to the lower $k-1$ rows of the classical ground state. It reads
\be
	\fL_{\text{Berry}}^2 = \frac{ \epsilon_{\mu\nu}}{2\lambda(\lambda+1)} \sum_{\alpha=2}^{k} \sum_{j=1}^\lambda\sum_{t=1}^{\lambda+1} 
	\Big[ p_\alpha (j-1 + (t-1)2\lambda) + p_{k+2-\alpha}(j-1+\lambda + (t-1)2\lambda)\Big]\partial_\mu\vec{\phi}^{\alpha,j,*} \cdot\partial_\nu\vec{\phi}^{\alpha,j}.
\ee	
The associated topological angles are 
\be
	\theta_{\alpha,j} = \frac{\pi}{\lambda(\lambda+1)}\sum_{t=1}^{\lambda+1} \Big[ p_\alpha(j-1 + 2\lambda(t-1))
	+ p_{k+2-\alpha}(j-1 +\lambda + (t-1)2\lambda)\Big]
\ee
for $\alpha=2,\cdots, k$ and $j = 1,\cdots, \lambda$. Again, the angles simplify to
\be
	\theta_{\alpha,j} = \frac{\pi}{\lambda} (p_\alpha+p_{k+2-\alpha})(j-1) + \pi p_{\alpha}
	\hspace{5mm}
	\alpha=2,\cdots, k; j=1,\cdots,\lambda
\ee
where we used the fact that $\lambda \pi p_\alpha \equiv 0$ since $\lambda$ is even. The correspondence between angle and condition on the $p_\alpha$ is provided in the following table.
\begin{table}[!h]
\centering
\begin{tabular}{|c |c |c c|} \hline
Subcase & Condition & Angles & \\ \hline \hline
$k$ even & 
\parbox{2cm}{\begin{align*} p_\alpha + p_{k+2-\alpha}  &= p_2 + p_k \\
(\lambda+1)(p_2+p_k) &= 2\lambda p_1 \\
\end{align*} } \vspace{-3mm} & \parbox{2cm}{\begin{align*} \theta_{\alpha,j}, &\theta_{k+2-\alpha,j} \\
&\theta_t \\
\end{align*} } \vspace{-3mm} & \hspace{10mm} \parbox{2cm}{\begin{align*} & \alpha=3,\cdots, \frac{k+1}{2}; \\
& j=1,\cdots,\lambda;  \hspace{3mm} t=1,\cdots,\lambda+1 \\
\end{align*} }\vspace{-3mm}
 \\ \hline
$k$ odd & \parbox{2cm}{\begin{align*} p_\alpha + p_{k+2-\alpha}  &= p_2 + p_k \\
2p_{\frac{k+2}{2}} &= p_2 + p_k \\
(\lambda+1)(p_2+p_k) &= 2\lambda p_1 \\
\end{align*} } & \parbox{2cm}{\begin{align*} \theta_{\alpha,j}, &\theta_{k+2-\alpha,j} \\
&\theta_{\frac{k+2}{2},j} \\
&\theta_t \\
\end{align*} }  &\hspace{10mm} \parbox{2cm}{\begin{align*}  & \alpha=3,\cdots, \frac{k}{2}; \\ &
j=1,\cdots,\lambda \\
&  t=1,\cdots,\lambda+1\\
\end{align*} } \\ \hline
\end{tabular} 
\caption{}
\end{table}

\item Case 5b: $n=k\lambda +1$ with $k>2$ and $\lambda>1$ with $\lambda$ odd. \newline Now that $\lambda$ is odd, the order of the unit cell has changed to  $\lambda(\lambda+1)$. The two parts of $\fL_{\text{Berry}}$ from Case 5a are modified to 
\be
	\fL_{\text{Berry}}^1 \to \epsilon_{\mu\nu} \frac{p_1}{\lambda(\lambda+1)} \sum_{t=1}^{\lambda+1} \sum_{j=1}^{\lambda}(t-1 +  (j-1)(\lambda+1))\partial_\mu\vec{\phi}^{t,*} \cdot\partial_\nu\vec{\phi}^t.
\ee
and
\be
	\fL_{\text{Berry}}^2 \to \frac{ \epsilon_{\mu\nu}}{\lambda(\lambda+1)} \sum_{\alpha=2}^{k} \sum_{j=1}^\lambda\sum_{t=1}^{\frac{\lambda+1}{2}} 
	\Big[ p_\alpha (j-1 + (t-1)2\lambda) + p_{k+2-\alpha}(j-1+\lambda + (t-1)2\lambda)\Big]\partial_\mu\vec{\phi}^{\alpha,j,*} \cdot\partial_\nu\vec{\phi}^{\alpha,j}
\ee
The angles are then
\be
	\theta_t = \frac{2\pi p_1}{(\lambda+1)}(t-1)
 \hspace{10mm}
	t=1,2,\cdots,\lambda+1.
\ee
and
\be
	\theta_{\alpha,j} = \frac{\pi}{\lambda} (p_\alpha+p_{k+2-\alpha})\left(j-1 + \frac{\lambda(\lambda-1)}{2}\right) + \pi p_{\alpha}
	\hspace{5mm}
	\alpha=2,\cdots, k; j=1,\cdots,\lambda
\ee
The same correspondence between condition and angle found in Case 5a applies here as well, so long as the slightly modified conditions for $\lambda$ odd are used (which can be found in Table~\ref{result2}).

\item Case 6: $n=k\lambda +c$ with $c\not=1,k-1$ \newline
In this case, the ground state order is $2\lambda(\lambda+1)$. Again, we split the Berry phase contribution into two pieces. The first $c$ rows contribute the following term:
\be
	\fL_{\text{Berry}}^1 = \frac{\epsilon_{\mu\nu}}{2\lambda(\lambda+1)} \sum_{j=1}^{\lambda} \sum_{t=1}^{\lambda+1} \sum_{\alpha=1}^c\Big[
	p_\alpha(t-1 + (j-1)2(\lambda+1)) + p_{c-\alpha+1}(t-1 + \lambda+1 + (j-1)2(\lambda+1))\Big] \partial_\mu \vec{\phi}^{t,\alpha,*}\cdot\partial_\nu\vec{\phi}^{t,\alpha}
\ee
which gives the topological angles
\be
	\theta_{t,\alpha} = \frac{\pi(p_\alpha + p_{c-\alpha+1})}{(\lambda+1)}(t-1)
+ \pi \lambda p_{c-\alpha+1}
+ \pi (\lambda-1)p_\alpha 
\hspace{10mm}
\alpha=1,\cdots, c; t=1,\cdots,\lambda+1
\ee
Meanwhile, the remaining $k-c$ rows contribute the term
\be
	\fL_{\text{Berry}}^2 = \frac{\epsilon_{\mu\nu}}{2\lambda(\lambda+1)}
	\sum_{j=1}^\lambda \sum_{t=1}^{\lambda+1}\sum_{\beta=c+1}^k
	\Big[ p_\beta(  j-1 + (t-1)2\lambda) + p_{k-\beta+c+1}( j-1 + (t-1)2\lambda + \lambda)\Big]  \partial_\mu \vec{\tilde\phi}^{j,\beta,*}\cdot\partial_\nu\vec{\tilde\phi}^{j,\beta}
\ee
which gives the topological angles
\be
	\tilde \theta_{j,\beta} = \frac{\pi(p_\beta + p_{k-\beta+c+1})}{\lambda}
(j-1)
+ \pi (\lambda+1) p_{k-\beta+c+1}
	 + \pi \lambda p_\beta
	 \hspace{10mm}
	 \beta=c+1,\cdots,k ; j=1,\cdots, \lambda.
\ee
Here, $\tilde\theta$ and $\vec\tilde \phi$ have been used in order to differentiate the two families of fields and topological angles, so as to simplify our notation. The relationships between topological angle and condition on the $p_\alpha$ topological angles are given in Table~\ref{tab6a}. 
\begin{table}[h]
\centering
\begin{tabular}{|c |c |c c|} \hline
Subcase & Condition & Angles & \\ \hline \hline
$k,c$ even & \parbox{2cm}{\begin{align*} 
p_\alpha + p_{c+1-\alpha}  &= p_1 + p_c \\
(\lambda+1)(p_\beta + p_{k-\beta +c+1}) &= \lambda (p_1+p_c)\\
\end{align*} } \vspace{-3mm}
 & \parbox{2cm}{\begin{align*} & \theta_{t,\alpha}, \theta_{t,c-\alpha+1}\\
&\tilde \theta_{j,\beta},\tilde\theta_{j,k-\beta+c+1} \\
\end{align*} } \vspace{-3mm} & \hspace{10mm} \parbox{2cm}{\begin{align*}  & \alpha=2,\cdots, \frac{c}{2}; t=1,\cdots, \lambda+1 \\
& \beta-c=1,\cdots,\frac{k-c}{2}; j=1,\cdots,\lambda \\
\end{align*} }\vspace{-3mm}
 \\ \hline
\parbox{2cm}{\begin{align*} k \text{ odd} \\
  c \text{ even} \\
  \end{align*}}
   & 
 \parbox{2cm}{\begin{align*} 
p_\alpha + p_{c+1-\alpha}  &= p_1 + p_c \\
(\lambda+1)(p_\beta + p_{k-\beta +c+1}) &= \lambda (p_1+p_c)\\
2(\lambda+1)p_{\frac{k+c+1}{2}} &= \lambda(p_1+p_c) \\
\end{align*} } \vspace{-3mm}
 & \parbox{2cm}{\begin{align*} & \theta_{t,\alpha}, \theta_{t,c-\alpha+1}\\
&\tilde \theta_{j,\beta},\tilde \theta_{j,k-\beta+c+1} \\
& \tilde \theta_{j, \frac{k+c+1}{2}} \\
\end{align*} } \vspace{-3mm} & \hspace{10mm} \parbox{2cm}{\begin{align*}  & \alpha=2,\cdots, \frac{c}{2}; t=1,\cdots, \lambda+1 \\
& \beta-c=1,\cdots,\frac{k-c-1}{2} \\
&  j=1,\cdots,\lambda \\
\end{align*} }\vspace{-3mm}\\ \hline
$k,c$ odd & \parbox{2cm}{\begin{align*} 
p_\alpha + p_{c+1-\alpha}  &= p_1 + p_c \\
2p_{\frac{c+1}{2}} &= p_1+p_c \\
(\lambda+1)(p_\beta + p_{k-\beta +c+1}) &= \lambda (p_1+p_c)\\
\end{align*} } \vspace{-3mm}
 & \parbox{2cm}{\begin{align*} & \theta_{t,\alpha}, \theta_{t,c-\alpha+1}\\
 & \theta_{t,\frac{c+1}{2}} \\
&\tilde \theta_{j,\beta},\tilde \theta_{j,k-\beta+c+1} \\
\end{align*} } \vspace{-3mm} & \hspace{10mm} \parbox{2cm}{\begin{align*}  & \alpha=2,\cdots, \frac{c-1}{2}; t=1,\cdots, \lambda+1 \\
& \beta-c=1,\cdots,\frac{k-c-1}{2} \\
&  j=1,\cdots,\lambda \\
\end{align*} }\vspace{-3mm}  \\ \hline
\parbox{2cm}{\begin{align*} c \text{ odd} \\
  k \text{ even} \\
  \end{align*}} & \parbox{2cm}{\begin{align*} 
p_\alpha + p_{c+1-\alpha}  &= p_1 + p_c \\
2p_{\frac{c+1}{2}} &= p_1+p_c \\
(\lambda+1)(p_\beta + p_{k-\beta +c+1}) &= \lambda (p_1+p_c)\\
2(\lambda+1)p_{\frac{k+c+1}{2}} &= \lambda(p_1+p_c) \\
\end{align*} } 
 & \parbox{2cm}{\begin{align*} & \theta_{t,\alpha}, \theta_{t,c-\alpha+1}\\
 & \theta_{t,\frac{c+1}{2}} \\
&\tilde \theta_{j,\beta},\tilde \theta_{j,k-\beta+c+1} \\
& \tilde\theta_{j,\frac{k+c+1}{2}} \\
\end{align*} } & \hspace{10mm} \parbox{2cm}{\begin{align*}  & \alpha=2,\cdots, \frac{c-1}{2}; t=1,\cdots, \lambda+1 \\
& \beta-c=1,\cdots,\frac{k-c-1}{2} \\
&  j=1,\cdots,\lambda \\
\end{align*} }  \\ \hline
\end{tabular} 
\caption{}
\label{tab6a}
\end{table}

\item Case 7a: $n=k\lambda + (k-1)$ with $\lambda$ odd \newline

The ground state has $2\lambda(\lambda+1)$-site order. The first $k-1$ rows have the following Berry phase contribution:
\be
	\fL^1_{\text{Berry}} = \frac{\epsilon_{\mu\nu}}{2\lambda(\lambda+1)}\sum_{\alpha=1}^{k-1}\sum_{j=1}^\lambda\sum_{t=1}^{\lambda+1} \Big[ p_\alpha ( t-1 + (j-1)2(\lambda+1)) + p_{k-\alpha}( t-1 + \lambda+1 + (j-1)2(\lambda+1))\Big] \partial_\mu\vec{\phi}^{\alpha,t,*}\cdot\partial_\nu\vec{\phi}^{\alpha,t}
\ee
The corresponding topological angles are
\be
	\theta_{t,\alpha} = \frac{\pi(p_\alpha + p_{k-\alpha})}{(\lambda+1)}  (t-1)
	+\pi p_{k-\alpha}
\ee

Meanwhile, the last row of the ground state contributes the term
\be
	\fL^2_{\text{Berry}} =  \frac{\epsilon_{\mu\nu}}{2\lambda(\lambda+1)}p_k\sum_{j=1}^{\lambda}\sum_{t=1}^{2(\lambda+1)} \left[ (j-1) + (t-1)\lambda\right]\partial_\mu\vec{\phi}^{j,*}\cdot\partial_\nu\vec{\phi}^j
\ee
giving rise to the angles
\be
	\theta_j = \frac{2\pi p_k}{\lambda}(j-1) + p_k \pi.
\ee
The correspondence between condition on the $p_\alpha$ and topological angle can be found in Table~\ref{taboo}.
\begin{table}[!h]
\centering
\begin{tabular}{|c |c |c c|} \hline
Subcase & Condition & Angles & \\ \hline \hline
$k$ odd & 
\parbox{2cm}{\begin{align*} p_\alpha + p_{k-\alpha}  &= p_1 + p_{k-1} \\
(\lambda+1)p_k &= \lambda( p_1+p_{k-1}) \\
\end{align*} } \vspace{-3mm} & \parbox{2cm}{\begin{align*} \theta_{\alpha,t}, &\theta_{k-\alpha,t} \\
&\theta_j \\
\end{align*} } \vspace{-3mm} & \hspace{10mm} \parbox{2cm}{\begin{align*} & \alpha=2,\cdots, \frac{k-1}{2}; \\
& j=1,\cdots,\lambda;  \hspace{3mm} t=1,\cdots,\lambda+1 \\
\end{align*} }\vspace{-3mm}
 \\ \hline
$k$ even & \parbox{2cm}{\begin{align*} p_\alpha + p_{k-\alpha}  &= p_1 + p_{k-1} \\
2p_{\frac{k}{2}} &= p_1 + p_{k-1} \\
(\lambda+1)p_k &= \lambda (p_1+p_{k-1}) \\
\end{align*} } & \parbox{2cm}{\begin{align*} \theta_{\alpha,t}, &\theta_{k-\alpha, t} \\
& \theta_{\frac{k}{2},t} \\
&\theta_j \\
\end{align*} }  & \hspace{10mm} \parbox{2cm}{\begin{align*}  & \alpha=2,\cdots, \frac{k-2}{2}; \\ 
& t=1,\cdots,\lambda+1 \\
& j=1,\cdots,\lambda\\
\end{align*} } \\ \hline
\end{tabular} 
\caption{}
\label{taboo}
\end{table}

\item Case 7b: $n=k\lambda + (k-1)$ with $\lambda$ even \newline
This is the final case. Now that $\lambda$ is even, the order of the unit cell has changed to $\lambda(\lambda+1)$. The two parts of $\fL_{\text{Berry}}$ are modified to 
\be
	\fL^1_{\text{Berry}} \to \frac{\epsilon_{\mu\nu}}{\lambda(\lambda+1)}\sum_{\alpha=1}^{k-1}\sum_{j=1}^{\frac{\lambda}{2}}\sum_{t=1}^{\lambda+1} \Big[ p_\alpha ( t-1 + (j-1)2(\lambda+1)) + p_{k-\alpha-2}( t-1 + \lambda+1 + (j-1)2(\lambda+1))\Big] \partial_\mu\vec{\phi}^{\alpha,t,*}\cdot\partial_\nu\vec{\phi}^{\alpha,t}
\ee
and
\be
	\fL^2_{\text{Berry}} \to  \frac{\epsilon_{\mu\nu}}{\lambda(\lambda+1)}p_k\sum_{j=1}^{\lambda} \sum_{t=1}^{(\lambda+1)} \left[ (j-1) + (t-1)\lambda\right]\partial_\mu\vec{\phi}^{j,*}\cdot\partial_\nu\vec{\phi}^j
\ee
and the angles are modified to
\be
	\theta_{\alpha, t} \to \frac{ \pi(p_\alpha + p_{k-\alpha})}{(\lambda+1)} (t-1) + \pi p_{k-\alpha} 
	+ \frac{\pi (p_\alpha + p_{k-\alpha})}{2}(\lambda-2)
\ee
and
\be
	\theta_j \to \frac{2\pi p_k }{\lambda}(j-1).
\ee

The correspondence between angles and conditions follows the same pattern as the previous case (7a), with a slight modification of the conditions themselves, according to Table~\ref{result3}.

\end{itemize}

\end{document}